\documentclass[sn-mathphys-num]{sn-jnl}

\usepackage{graphicx}%
\usepackage{multirow}%
\usepackage{amsmath,amssymb,amsfonts}%
\usepackage{amsthm}%
\usepackage{mathrsfs}%
\usepackage[title]{appendix}%
\usepackage{xcolor}%
\usepackage{textcomp}%
\usepackage{manyfoot}%
\usepackage{booktabs}%
\usepackage{algorithm}%
\usepackage{algorithmicx}%
\usepackage{algpseudocode}%
\usepackage{listings}%

\theoremstyle{thmstyleone}%
\theoremstyle{thmstyletwo}%

\theoremstyle{thmstylethree}%

\begin{document}

\title{Experimental observation of gapped shear waves and liquid-like to gas-like dynamical crossover in active granular matter}


\author[1,2,3,+]{Cunyuan Jiang}
\author[1,4,+]{Zihan Zheng}
\author[1,4]{Yangrui Chen}
\author[1,2,3,*]{Matteo Baggioli}
\author[1,2,3,*]{Jie Zhang}
\affil[1]{School of Physics and Astronomy, Shanghai Jiao Tong University, Shanghai 200240, China}
\affil[2]{Wilczek Quantum Center, Shanghai Jiao Tong University, Shanghai 200240, China}
\affil[3]{Shanghai Research Center for Quantum Sciences, Shanghai 201315, China}
\affil[4]{Institute of Natural Sciences, Shanghai Jiao Tong University, Shanghai 200240, China}

\affil[*]{Corresponding authors: \color{blue}b.matteo@sjtu.edu.cn\color{black}, \color{blue}jiezhang2012@sjtu.edu.cn \color{black}}

\affil[+]{These authors contributed equally to this work}


\abstract{Unlike crystalline solids, liquids lack long-range order, resulting in diffusive shear fluctuations rather than propagating waves. Simulations predict that liquids exhibit a $k$-gap in wave-vector space, where solid-like transverse waves reappear above this gap. Experimental evidence in classical liquids has been limited, observed only in 2D dusty plasmas. Here, we investigate this phenomenon using active Brownian vibrators and uncover distinct gas-like and liquid-like phases depending on the packing fraction. We measure key properties, including pair correlation functions, mean square displacements, velocity auto-correlation functions, and vibrational density of states. In the liquid-like phase, we confirm the $k$-gap in transverse excitations, whose size grows as the packing fraction decreases and eventually disappears in the gas phase. Our findings extend the concept of the $k$-gap to active granular systems and reveal striking parallels with supercritical fluids.}




\maketitle

\section*{Introduction}
Collective modes are a direct macroscopic manifestation of coherent atomic motion and have a pivotal role in determining the thermodynamic, mechanical, and transport properties of physical systems. Phonons, collective lattice vibrations in solids, constitute an emblematic example as they determine most of the physics of solids at low energy, including their density of states, their heat capacity (Debye theory), and even possible superconducting instabilities (BCS theory). Phonon dynamics can be described using elasticity theory \cite{chaikin_lubensky_1995} or hydrodynamics \cite{PhysRevA.6.2401}, from which one derives that their frequency at long-wavelength is linear in the wavevector $k$, $\omega_{L,T}=v_{L,T} k$, with the transverse (T) and longitudinal (L) speeds of sound governed by the elastic moduli. 

Because of the random atomic distribution and the absence of a fixed equilibrium reference frame, the fate of phonons and the vibrational properties of liquids represent a much harder challenge for both theory and experiments \cite{Trachenko_2016}. In liquids, the dynamics of longitudinal long-wavelength fluctuations have been experimentally ascertained \cite{Pilgrim_2006} to be qualitatively identical to that of solids, despite a smaller sound speed. On the contrary, the dynamics of transverse (or shear) long-wavelength fluctuations are radically different. Liquids have a vanishing static shear modulus and, at small wave-vector, they display a shear diffusion mode rather than propagating shear waves (transverse phonons) as in solids \cite{chaikin_lubensky_1995}. 

Leveraging on a simple viscoelastic model, Maxwell \cite{maxwell1867iv} proposed that shear stress in liquids has a characteristic exponential decay time $\tau_M \equiv \eta/G_\infty$, where $\eta$ is the shear viscosity and the instantaneous shear modulus $G_\infty$. This timescale is now known as the Maxwell relaxation time. Based on a more microscopic picture of liquid dynamics, Frenkel later proposed \cite{Frenkel1946} to identify such a timescale with the time of local particle re-arrangements, corresponding to hopping processes over potential barriers.

The emerging Maxwell-Frenkel picture of liquid dynamics (see \cite{trachenko2023theory} for the complete history lesson) suggest that shear fluctuations in liquids obey the following telegrapher equation,
\begin{equation}
    \omega_T^2+i \omega_T \tau_M^{-1}= v_T^2 k^2. \label{tele}
\end{equation}
In Eq.~\eqref{tele}, $v_T$ is the transverse speed of sound related to the instantaneous shear modulus $G_\infty$. By solving \eqref{tele} (see \cite{BAGGIOLI20201} for an extensive review), the dispersion of shear waves is obtained,
\begin{equation}
    \omega_T=-\frac{i}{2 \tau}+ v_T\sqrt{ k^2-k_g^2}\qquad \text{with}\qquad k_g^{-1}\equiv  2 v_T \tau.
\end{equation}

For long-wavelengths, one recovers the hydrodynamic shear diffusion mode with collective diffusion constant $D_c \equiv v^2_T \tau = \eta/\rho$ (with $\rho$ the mass density of the system) predicted by Navier-Stokes equations \cite{landau2013fluid}.
Above a critical wave-vector $k_g$, known as $k$-gap, the real part of the frequency becomes nonzero. Above $k_g$, when the real part of $\omega_T$ becomes larger than its imaginary part, propagating solid-like shear waves are then expected to emerge in liquids. This leads to a corresponding elastic-like response below a certain critical distance, $L_c \equiv 2\pi/k_g$. In simpler terms, it means that liquids are predicted to exhibit solid-like transverse vibrational modes not only for high-frequency, $\omega \gg 1/\tau$ (as proposed initially by Frenkel \cite{Frenkel1946} based on a single particle picture), but also for large wave-vectors $k \gg k_g$.

The $k$-gap is expected to appear at the melting temperature and to expand into the liquid phase as the temperature increases \cite{PhysRevLett.118.215502}. It then reaches the maximum wave vector allowed at the edge with the gas phase. The existence of the $k$-gap, and its properties as described by equation \eqref{tele}, have been confirmed by several molecular simulations of classical liquids \cite{PhysRevLett.118.215502,PhysRevE.107.014139,Fomin_2020,Kryuchkov2019,TOLEDOMARIN2019100030,10.1063/1.5050708,bryk2023propagation,10.1063/1.4997640} and other liquid systems \cite{10.1063/5.0054854,10.1063/1.5088141,PhysRevE.107.055211}, indicating the validity of the theoretical framework, even from a quantitative perspective \cite{PhysRevB.101.214312}. Additionally, the similarity between vibrational modes at high frequency/wave-vector in liquids and solids \cite{Ruocco1996,doi:10.1073/pnas.1006319107,RevModPhys.77.881}, and the solid-like nature of confined liquids at low frequencies, as predicted by the concept of an elastic critical length $L_c$, have been experimentally verified \cite{Noirez_2012,yu2023unveiling}. However, due to the limitations of experimental scattering techniques at low $k$ and $\omega$, the $k$-gap has not been observed experimentally in classical liquids. The only recorded observation of the $k$-gap has been in complex dusty plasmas \cite{PhysRevLett.97.115001}, which are systems of charged particles that interact strongly via Coulomb forces and exhibit liquid-like collective dynamics \cite{10.1063/5.0168088}.

Granular materials differ from conventional thermal equilibrium systems, such as molecular gases, liquids, solids, colloidal liquids, or solids. They are made up of macroscopic particles that experience negligible thermal fluctuations compared to the typical energy scales of the system. Granular materials also have high dissipation due to inter-particle solid friction in the dense solid-like phase or inter-particle inelastic collisions in the fluid-like phase \cite{Behringer-RMP}. This means that a continuous external energy injection is needed to maintain the fluid-like phase of a granular system, making it a prototype of systems far from thermal equilibrium. However, this raises questions about whether the $k$-gap description, commonly used for molecular liquids or plasma, applies to a granular fluid. More in general, it remains unclear whether collective modes of a granular fluid are similar to those of classical liquids \cite{Trachenko_2016} and, if so, what plays the role of thermodynamic variables such as temperature.

The densely packed configurations of granular materials often exhibit fluid-like behaviors when subjected to external forces \cite{Howell-Behringer-PRL-1999, Midi-EPJE, Corwin-Nature-05, Pouliquen-Nature-06, vanHecke-PRL-2010, Kou2017}. However, due to the presence of permanent contacts and force chains \cite{Behringer-Nature-05}, measuring the Hessian matrix directly in densely packed granular matter has proven to be a challenging task, even in the quasi-static limit \cite{Zhangling-NC, Wangyinqiao-PRB, Zhangling-PRR-2021, Zhangling-PRB-2021}.
In contrast, loose granular matter, where collisions primarily govern particle interactions, has seen theoretical analyses of hydrodynamics and collective modes in granular fluids. This involves formulating transport equations for essential hydrodynamic quantities like mass, momentum, and heat, followed by a linear stability analysis of the homogeneous states \cite{10.1063/1.858716, PhysRevE.58.4638, Noije-PRE-99, PhysRevE.77.031310, Puglisi-JSM-11, PhysRevE.83.011301, Soto-PRE-13}. While transverse modes decouple \cite{Puglisi-JSM-11}, the longitudinal sector becomes intricate due to the non-conservation of energy, leading to significant modifications in the longitudinal channel. To the best of our knowledge, the discussion of the $k$-gap in the hydrodynamics of granular fluids has been absent, as it extends beyond the conventional long-wavelength hydrodynamic description. From an experimental standpoint, the primary advantage of granular fluids over traditional molecular liquids is the macroscopic size of the particles, which are measured in centimeters in this case. This larger size significantly facilitates the tracking of particle positions and dynamics using cameras. In contrast, such tracking is extremely challenging, if not impossible, with molecular liquids. In those cases, excitations can only be investigated through techniques like X-ray or inelastic neutron scattering, which are far more complicated, especially in the frequency and wave-vector ranges where the $k$-gap is expected to emerge.

\begin{figure}[htbp]
    \centering
    \includegraphics[width=0.65\linewidth]{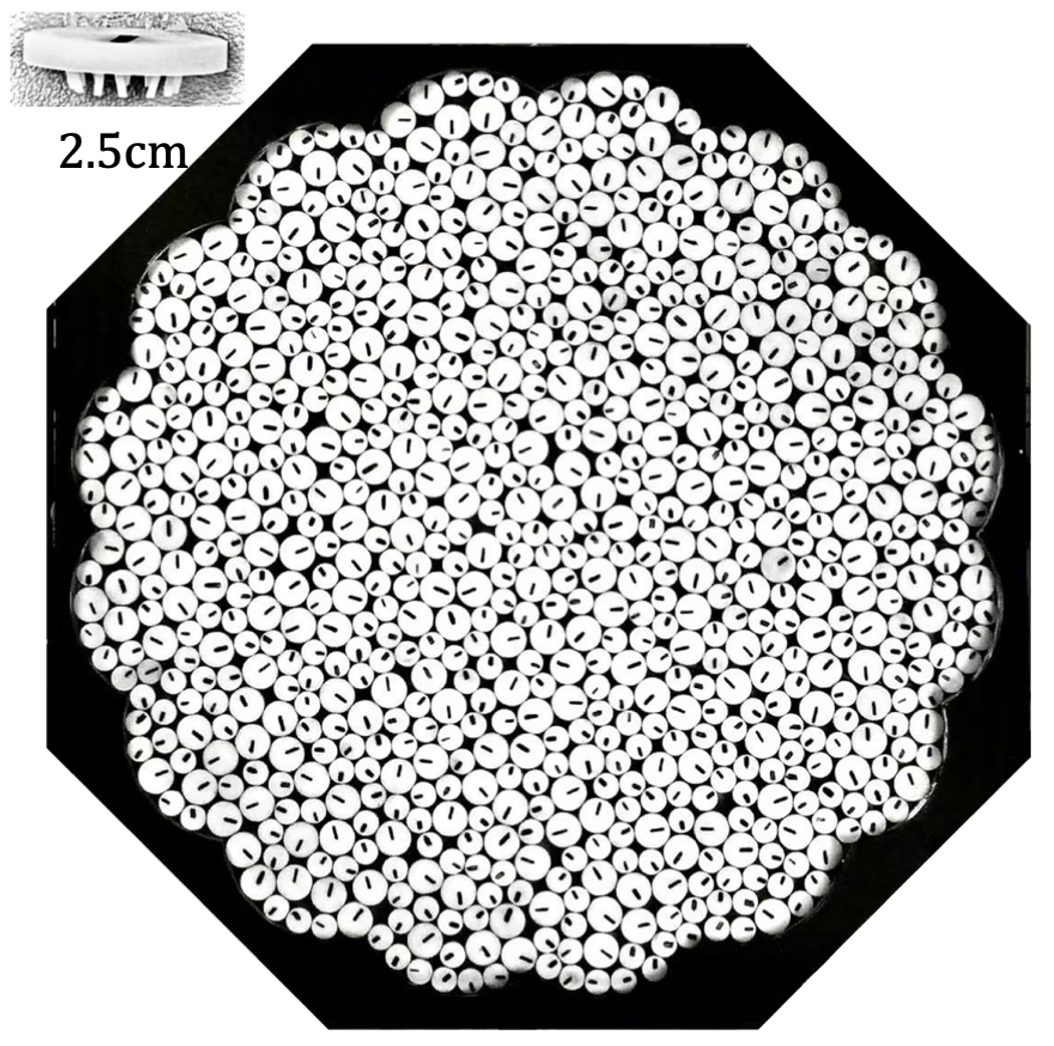}
    \caption{\textbf{Active granular matter in the Lab.} Top view of a layer of active Brownian vibrators. The packing fraction in this layer is $\phi=0.822$. In the upper left corner, one Brownian vibrator is shown. The total area of large and small particles has a fixed ratio of $1/1$, and their diameters have a ratio of $d_l / d_s = 1.4/1$.}
    \label{fig:0}
\end{figure}

If the k-gap description can apply to granular fluids, then a homogeneously driven granular fluid would be the most straightforward scenario to explore. However, creating a homogeneously driven experimental system has proven challenging \cite{Poschel-PRl, Yangrui-PRE, Yangrui-NC} due to the influence of gravity and anisotropic driving in three-dimensional (3D) systems or the implementation of boundary driving in quasi-two-dimensional (2D) vertical systems \cite{Menon-PRL2000, Sano-JFM-2009}. Additionally, the influence of 3D effects in quasi-2D horizontal systems has made it difficult to achieve homogeneous driving \cite{Olafsen-PRL-98, Losert-Chaos-99, Olafsen-PRE-02}. While a few quasi-2D systems have achieved homogeneous driving \cite{Shattuck-PhysRevLett.98.188301,Shattuck-PhysRevLett.96.258001,Poschel-PRl}, some lacked single-particle velocity Gaussian statistics \cite{Shattuck-PhysRevLett.98.188301,Shattuck-PhysRevLett.96.258001}. In contrast, others incorporated persistent unidirectional rotation at the single-particle level \cite{Poschel-PRl}, introducing additional complexities in energy injection at the single-particle scale. Recently, Chen et al. designed an experimental system that achieves homogeneous driving, single-particle velocity and rotation statistics with Gaussian distributions of zero means in a quasi-2D system \cite{Yangrui-PRE, Yangrui-NC}. This system closely aligns with the active Brownian particles introduced in theoretical studies from the perspective of active matter \cite{CapriniPRL-2023}. We notice that our experimental setup can be considered as an active system since the external energy input, that maintains the system out of equilibrium, acts individually and independently on each ``active particle'' \cite{PhysRevX.12.010501}.

Compared to a dusty plasma, a nonequilibrium system made up of micron-sized charged particles suspended in a plasma, an active granular system differs significantly in its interaction potential and driving mechanism. In dusty plasmas, the potential is governed by a Yukawa potential, while in active granular systems, interactions occur through inelastic collisions and solid friction. Additionally, the driving mechanisms differ: dusty plasma is driven by laser heating at the boundaries, while active granular systems experience homogeneous and random driving of individual particles. Our experimental findings add to the previous experimental observation of a $k$-gap dispersion in a dusty plasma \cite{PhysRevLett.97.115001}, demonstrating the universality of the $k$-gap phenomenon in the liquid phases of matter.

Active granular systems provide a novel platform for exploring the emergence of collective dynamics and showcasing a rich interplay of complex phases and phenomena. Our study focuses on bi-disperse active Brownian vibrators. Through measurements of the pair correlation functions, mean square displacements, velocity auto-correlation functions, vibrational density of states, and a detailed analysis of particle motion, we demonstrate that this active system exhibits both gas-like and liquid-like phases, depending on the packing fraction, despite pure hard-disk-like repulsive interactions. Within the granular liquid-like phase, we experimentally validate the existence of a $k$-gap in the dispersion of transverse excitations. This gap becomes more significant with a decrease in packing fraction and becomes ill-defined in the gas phase because of the disappearance of well-defined modes, aligning with theoretical expectations. Our results offer a direct experimental confirmation of the $k$-gap phenomenon, extending its relevance beyond classical thermal liquids to active granular systems, and reveal the existence of similarities between the physics of active granular matter and supercritical fluids.

\section*{Result and discussions}
\subsection*{Granular Brownian vibrators}
Fig.\ref{fig:0} displays the top view of a layer of bi-disperse Brownian vibrators positioned on the surface of a shaker. This layer of particles is confined within a flower-shaped boundary to prevent the creep motion of particles near the boundary. Each Brownian vibrator has a flat, disk-shaped cap with twelve alternatively inclined legs below the cap. When a vertical sinusoidal vibration is applied to the supporting base, each single Brownian vibrator performs 2D Brownian motion. Previous studies by Chen et al. revealed that the translational and rotational velocities of a single Brownian vibrator follow Gaussian distributions with zero means \cite{Yangrui-PRE}. 
These features lead us to term the particle an ``active Brownian particle'', closely mimicking the conditions studied in theoretical investigations of active matter systems \cite{CapriniPRL-2023}.
Moreover, in a collection of Brownian vibrators of the same size, the translational velocities of individual particles follow a Maxwell-Boltzmann distribution for low and intermediate speeds, but show high-energy tails that deviate significantly from the Maxwell-Boltzmann distribution for large speeds, which can be attributed to the inelastic collisions of particles and the homogeneous driving \cite{Yangrui-PRE}. Unlike the mono-disperse systems studied earlier \cite{Yangrui-PRE, Yangrui-NC}, the present system is bi-disperse, which prevents crystallization at high packing fractions, as depicted in Fig.\ref{fig:0}. More details about this setup are presented in the Methods. We emphasize that the frequency of the vertical sinusoidal vibration applied is of $100$ Hz. In the rest of the manuscript, we will consider only time-scales which are parameterically longer than the driving frequency. In that regime, equilibrium thermodynamic and hydrodynamic concepts can still be applied.

Unlike conventional polar particles studied previously \cite{Dauchot-PRL2010,Kumar2014,PhysRevLett.110.238301}, the lack of particle-scale built-in asymmetry makes our experimental system unique and novel. The active force on each nonpolar particle results from collisions between the tilted legs and the vibrating bottom surface, causing the particle's central axis to tilt slightly away from gravity, allowing only some legs—often represented as a single leg—to be propelled upon contact with the surface. These interactions produce minimal correlation in the contact angle, leading to random driving force directions while maintaining a nearly constant magnitude (see Refs. \cite{Yangrui-PRE, Yangrui-NC} for details).
In this system, there is no fluid flow from a surrounding solvent, and the main dissipation with the environment results from friction and inelastic collisions. Consequently, our system naturally belongs to the category of dry active matter.

In our present system of bi-disperse (nonpolar) Brownian vibrators, we have observed no global flocking within the several-hour experimental time window, which is likely due to the disorder introduced by bidispersity, in contrast to our previous monodisperse systems \cite{Yangrui-NC}, where we observed global flocking, aligning with the theoretical investigation of active Brownian particles \cite{CapriniPRL-2023}. Furthermore, we have not observed any phase separation, unlike the self-propelled binary colloids known as Quincke rollers, where significant demixing of small and large colloidal particles occurs after the system begins a global rotational collective motion \cite{Maity-PRL2023}. On a microscopic level, the interactions between these Quincke rollers are influenced by electrostatic and hydrodynamical forces, which are quite different from the interactions observed in our nonpolar granular disks in a dry environment. While examining phase separation in nonpolar disks is an promising area for future research, it is currently beyond the scope of this study. Real-time videos of the particle motion are provided with from the Supplementary Movie 1 to 4 for different packing fraction to confirm these statements.

\begin{figure}[ht!]
    \centering
    \includegraphics[width=\linewidth]{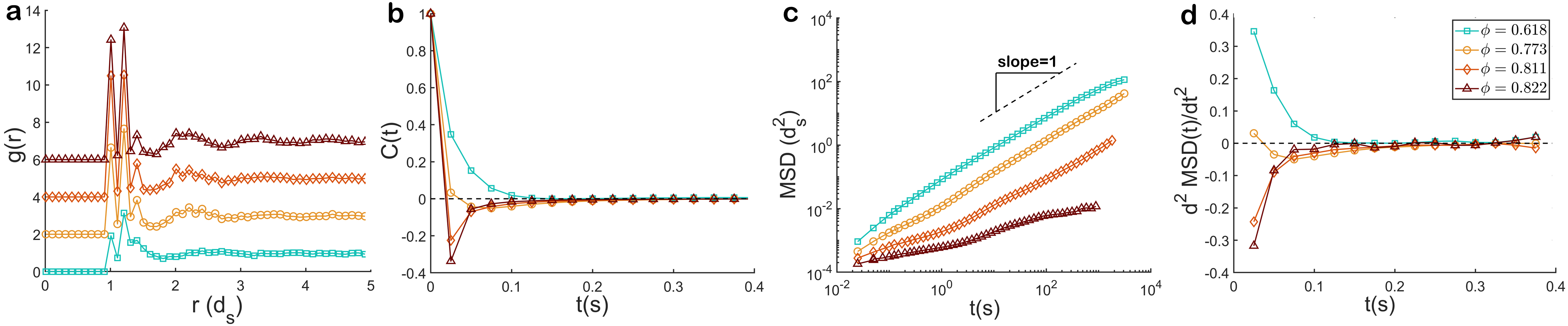}
    \caption{\textbf{Disappearance of medium-range order and the dynamical transition between a liquid-like to gas-like phase.} \textbf{(a)}: The pair distribution functions $g(r)$. \textbf{(b)}: The normalized velocity auto-correlation functions (VACF). \textbf{(c)}: The mean square displacement (MSD) of particles. \textbf{(d)}: The second derivative of MSD with respect to time. Here, the values of MSD are properly normalized using $2Z(0)$, where $Z(t)$ is defined in \eqref{eq-3}. In panels (a-d), the same set of packing fractions $\phi$ is chosen following the color scheme described in the legend in panel (d).}
    \label{fig:1}
\end{figure}

\subsection*{Structural and dynamical crossovers}
We experimentally investigate our active granular system by measuring the pair correlation function $g(r)$. In Panel (a) of Fig.\ref{fig:1}, we show the results for different values of packing fraction $\phi$, which is the fractional area occupied by the particles over the whole system. The first set of peaks corresponds to three peaks of $g(r)$ within the range of $1d_s\le r \le 2 d_s$. It arises from the bi-dispersity of particle sizes, causing a single peak to split into three peaks. The second peak of $g(r)$ is located within $2 d_s \le r \le 3 d_s$, and the third peak is in the range of $3 d_s \le r \le 4 d_s$. As we decrease the packing fraction, we observe a considerable decrease in the height of the first set of peaks, and the second and third peaks disappear. This observation implies that as we decrease the packing fraction, the medium-range order vanishes, and the system undergoes a structural crossover. This transition occurs at about $\phi=0.618$ and is known in the context of supercritical fluids as the Fisher-Widom line \cite{10.1063/1.1671624}. 

To establish a connection between structure and dynamics, as achieved for supercritical fluids in \cite{10.1063/1.4844135}, in panels (b)-(d) of Fig.~\ref{fig:1}, we present the experimental results for the velocity auto-correlation functions (VACF) $C(t)$, the mean square displacements (MSD), and their corresponding second derivatives with respect to time $\frac{d^2}{dt^2}\text{MSD}(t)$, as functions of time for various values of the packing fraction $\phi$.

We begin by defining the unnormalized VACF $Z(t)$ as
\begin{equation} \label{eq-3}
    Z(t) \equiv \langle v_i(0) v_i(t) \rangle = \frac{1}{d}\langle \vec{v}(0) \vec{v}(t) \rangle,  
\end{equation}
where the index $i$ specifies the Cartesian component of the velocity $\vec{v}$ with $i=x,y$ and $d=2$ for our system. The statistical average $\langle \cdot \rangle$ is first taken over different initial times `0' for a given particle and then over all particles. The VACF $C(t)$ is then defined as \( C(t) \equiv Z(t)/Z(0)\).

The VACF of a gas decreases continuously with time, while for liquids near melting point and solids, it shows a combination of an oscillatory and a decaying term. The presence of an oscillatory part in the $C(t)$  can be identified by looking for the occurrence of a minimum or a change in its slope. For high packing fractions, $C(t)$  displays a clear minimum below $t \approx 0.1 s$ that gradually disappears as the packing fraction decreases, as shown in Fig.\ref{fig:1}(b). At a packing fraction $\phi=0.618$, the minimum in the VACF is no longer present, and $C(t)$ becomes a continuously decreasing function, as expected in a gas. In the field of supercritical fluids, this dynamical transition determines the so-called Frenkel line that separates the rigid liquid phase, presenting oscillatory motion, and the non-rigid gas-like fluid phase. Evidence for the structural nature of the Frenkel line, hinting towards a possible equivalence with the Fisher-Widom line concept, has been reported in supercritical fluids \cite{PhysRevLett.111.145901}. Despite the complete equivalence between the structural and dynamical criteria remains unproved, a direct connection between the dynamical crossover and thermodynamics has been demonstrated \cite{Simeoni2010}. Aware of these distinctions, in the rest of the manuscript, we will adopt the jargon rigid (liquid-like) and non-rigid (gas-like) states interchangeably. According to Frenkel's theory \cite{Frenkel1946}, this dynamical crossover corresponds microscopically to a situation where the jumping time between oscillatory motion around different local minima of the potential becomes comparable with the shortest vibration time. This dynamical crossover is also expected to coincide with the disappearance of collective shear waves at all frequencies in the liquid. Frenkel idea relies on a single particle picture, while the $k$-gap equation \eqref{tele} describes collective dynamics. The time-scale $\tau$ appearing in Eq.~\eqref{tele} is therefore more correctly identified with the Maxwell relaxation time. In simple fluids the Maxwell time is very close to the lifetime of local connectivity \cite{PhysRevLett.110.205504}, that is another single particle concept that will be analyzed below. The critical packing fraction of $\phi=0.618$ is very close to the value at which medium-range order disappears in the pair correlation functions shown in panel (a). This suggests a significant link between structure and collective dynamics in active granular systems, similar in spirit to the results presented in \cite{PhysRevLett.111.145901}. We notice that there is now firm experimental evidence that the change of particle dynamics at the Frenkel line is seen in structural changes \cite{COCKRELL20211}. Despite a complete analysis of these structural changes being beyond the scope of the present work, in the Supplementary note 2, we provide a preliminary study of the experimental structure factor $S(k)$ for four different packing fractions that confirms the gradual disappearance of structural order by decreasing the packing fraction. The relation between structural changes and dynamics in liquids is still poorly understood (see, for example, \cite{trachenko2023theory}). It would be necessary to explore this connection further in granular fluids.

Another substantial dynamical quantity, besides the VACF, is the MSD of particle motion. This is shown in panel (c) in Fig.\ref{fig:1}, where curves for the same set of packing fractions $\phi$ as in panel (b) are drawn. For $\phi=0.618$ and below (data not shown), MSD is quasi-ballistic for $t<0.1 s$, indicating underdamped particle dynamics. For $t>0.1 s$, the slope of MSD is close to one, showing the diffusive motion of a particle. However, for very large $t$, the slope of MSD deviates from one due to the finite system size as the length scale of MSD gradually approaches the system size. For $\phi>0.618$, the quasi-ballistic motions at small $t$ are not clearly visible, and a subdiffusive regime at intermediate times emerges. The upper bound of this regime depends on the value of $\phi$, beyond which the diffusive behaviors recover for the curves corresponding to $\phi=0.773$ and $\phi=0.811$. At $\phi=0.822$, within the entire observation time window of $0.025 s \le t \le 1000 s$, the MSD shows subdiffusive behaviors, indicating the progressively more significant glassy dynamics with the increment of $\phi$.

The MSD and the VACF are closely related to each other since $\frac{d^2}{dt^2}\mathrm{MSD}(t)=4Z(t)$ \cite{hansen2013theory}. This equivalence has been already utilized in Ref.~\cite{PhysRevLett.111.145901} to investigate the ``liquid-gas'' transition. Upon comparing panels b and d of Fig.\ref{fig:1}, we experimentally verified this equivalence, and the results show an excellent quantitative mutual agreement.

\begin{figure}[ht!]
    \centering
    \includegraphics[width=0.65\linewidth]{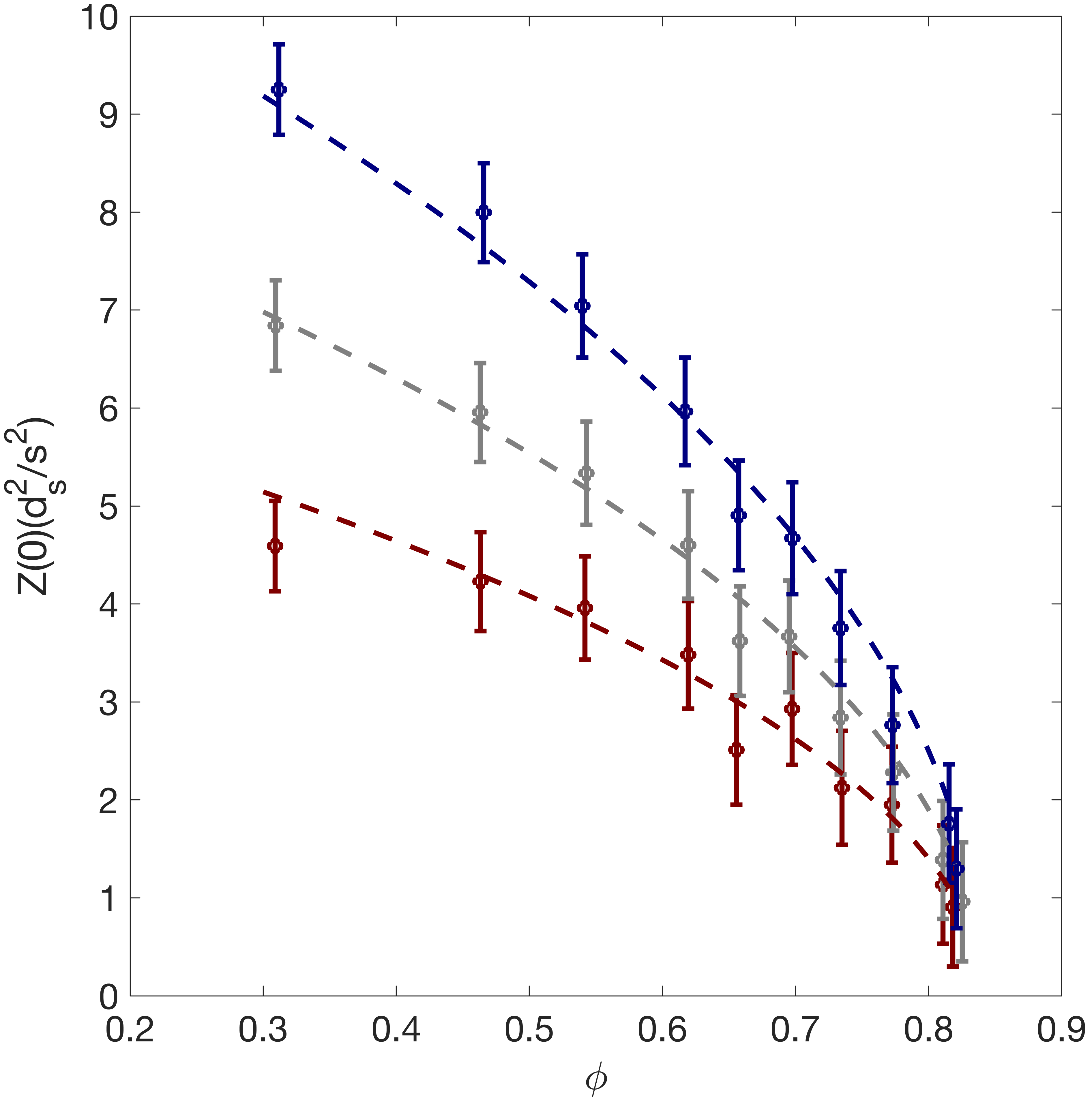}
    \caption{\textbf{The role of the packing fraction in active granular matter.} The average velocity squared $Z(0)$ for small (red), large (blue) and all (gray) particles as a function of the packing fraction $\phi$. All the data are well fitted by a phenomenological function $Z(0)=a \sqrt{b-\phi}$, where $b=0.8318$ for all curves. The error bar of averaged value here is obtained as the difference of minimum and averaged value.}
    \label{fig:temp}
\end{figure}

Granular systems are athermal in nature because of the macroscopic size of their constituents. At fixed activity, it is therefore necessary to understand which parameter plays an analogous role of temperature in thermal fluids, driving the system from a gas-like to a liquid-like phase. Both structural and dynamical observables suggest that the packing fraction $\phi$ of the Brownian vibrators plays such a role. In order to provide further qualitative evidence for this analogy, in Fig.\ref{fig:temp}, we plot the average velocity squared $Z(0)$, that is proportional to the average kinetic energy, as a function of the packing fraction for small (red), large (blue) and all (gray) particles. We observe a clear anti-correlation between the two quantities, which is well fitted by a phenomenological function $Z(0)\propto \sqrt{\phi_c-\phi}$, with $\phi_c \approx 0.8318$. In the literature (\textit{e.g.}, \cite{PhysRevE.101.032903}), $Z(0)$ has been often associated to an effective granular temperature, $T_e$. Our results demonstrate therefore that $T_e$ anti-correlates with $\phi$, consistent with the dynamics experimentally observed both at the particle-level and collective scale.
Temperature is clearly defined only in equilibrium thermodynamic systems. Defining temperature rigorously in non-equilibrium systems is often very challenging.  Here, we introduce the term `temperature’ for our system only in a vague, intuitive sense, drawing an analogy with well-defined meanings for ordinary materials made of molecules in thermal equilibrium.
It is necessary to notice that the scaling of the experimental data follows a mean-field behavior. We are not aware of any theoretical explanation of this phenomenon. This analysis suggests the existence of a critical packing fraction $\phi_c$ which might be connected to a jamming type transition in active granular systems. We leave the exploration of these two points for future research. Finally, we notice that the average velocity squared $Z(0)$ for large particles is consistently larger than that of small particles, with this difference becoming more pronounced for small packing fraction. Given that the steady state velocity is a result of the balance between the energy injected and the energy loss caused by friction, this might be explained by the fact that larger particles experience stronger drag force induced by activity.

\begin{figure}[ht!]
    \centering
    \includegraphics[width=\linewidth]{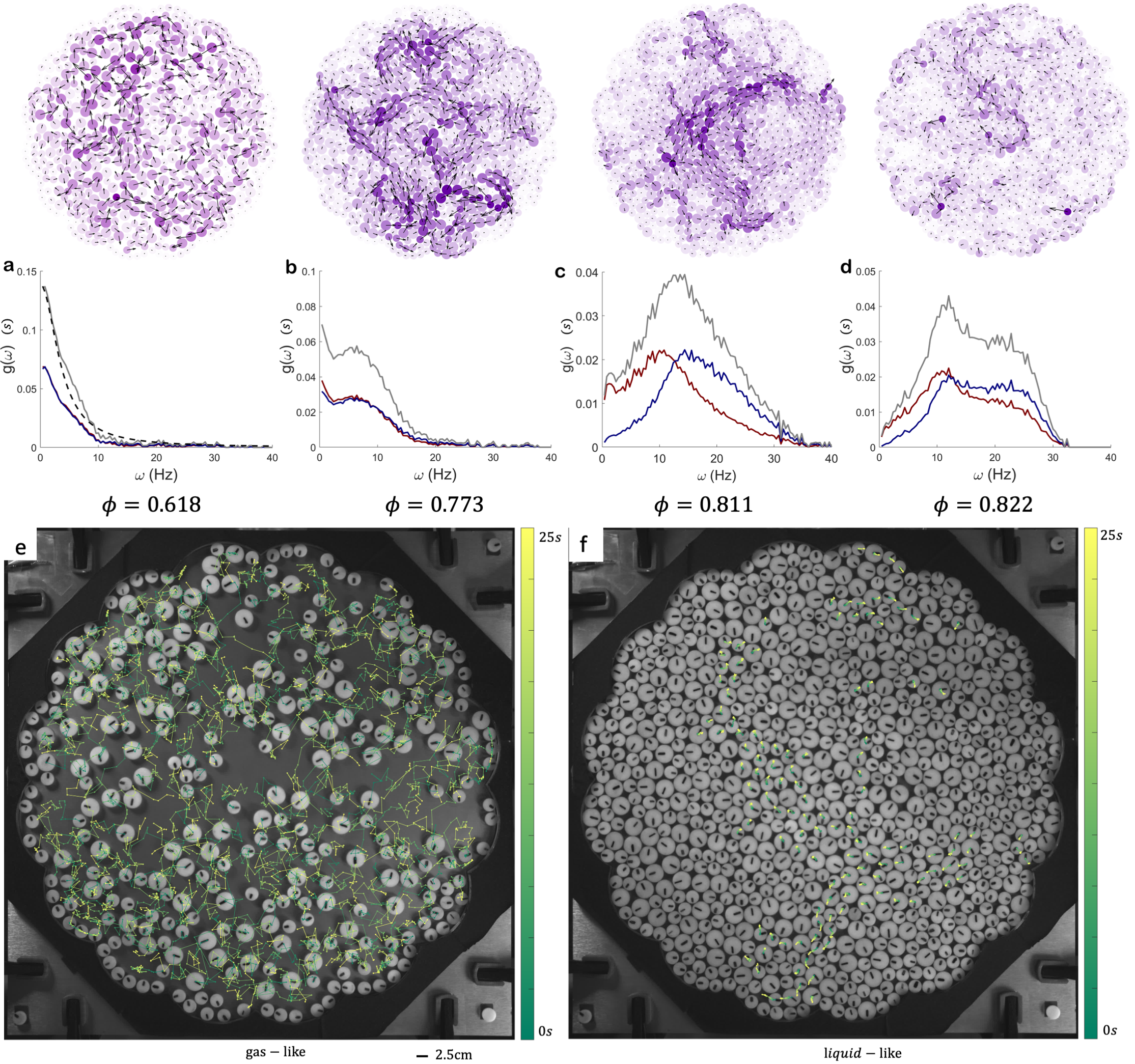}
    \caption{\textbf{From isotropic gas-like motion to collective liquid-like dynamics.} The experimental vibrational density of states (VDOS) for packing fraction $\phi=0.618$, $\phi=0.773$, $\phi=0.811$, and $\phi=0.822$ from panels (a) to (d). The VDOS is represented by gray, blue, and red lines, indicating the total VDOS and longitudinal and transverse components respectively. The VDOS is normalized by setting its area $\int g(\omega) d\omega$ to $1$, and the frequency is measured in Hz. The black dashed line in panel (a) shows the Lorentzian line shape $\alpha/(\omega^2+\alpha^2)$ with $\alpha=3.3$. The top insets show the displacement configuration, with the intensity of the color indicating the amplitude of the single particle displacement, with darker shades representing more significant displacement. The displacement vectors have been enlarged $\times 1$, $\times 2$, $\times 3$, and $\times 10$ in panels (a) to (d), respectively. The bottom panels (e) and (f) show the trajectories of each particle, with the color changing from green to yellow indicating time evolution.}
    \label{fig:2}
\end{figure}

The radial pair correlation function is defined as,
\begin{equation}
    g(r) = \dfrac{1}{2 \pi r N \rho} \sum_{i=1}^{N}\sum_{j\neq i}^{N} \delta\left(r-|\vec{r}_{ij}|\right),
\end{equation}
where $\vec{r}_{ij}$ is the vector between $i$th and $j$th particles, $N$ is the total number of particles, and $\rho$ is the particle density of system. As shown in Fig.\ref{fig:1} a, three peaks appear below $r= 2 d_s$. The first peak is at $r = 1 d_s$, which indicates small-small particle pairing, the second indicates small-big particle pairing at $r = 1.2 d_s$, and the third is big-big particle pairing at $r = 1.4 d_s$, which is also the diameter of the big particle.

\subsection*{From gas-like to liquid-like dynamics}
To study collective motion, we examine displacement vectors and vibrational density of states (VDOS) $g(\omega)$, which can be obtained by diagonalizing the dynamical matrix computed from the displacement correlation matrix (see Methods). 

In Fig.\ref{fig:2}, panel (a), we observe that for the lowest packing fraction data, the VDOS decreases monotonically with frequency and can be accurately described by a Lorentzian line shape, $g(\omega)=\alpha/(\alpha^2+\omega^2)$ (dashed black line), at least for frequencies below 5 Hz. This line shape is indicative of purely Langevin diffusive dynamics. It is typical of a gas-like state, where particle collisions are almost uncorrelated and independent and can be described by kinetic theory. The Lorentzian fit becomes less accurate at higher frequencies, indicating that the low-packing fraction system is a dilute liquid rather than an ideal gas of free particles. Additionally, the longitudinal and transverse components of $g(\omega)$ are identical, confirming the emergent isotropy of the low-packing fraction phase. Due to the dilute packing, the constituent particles exhibit random, uncorrelated motion in both amplitude and direction. Individual particles' motion is collisional, resulting in substantial displacements away from their initial positions.

Upon increasing the packing fraction and entering the rigid liquid phase described above, the VDOS undergoes significant changes. A weak and broad peak emerges around 7 Hz in both the transverse and longitudinal components, indicating the emergence of strongly overdamped collective motion. The VDOS is no longer monotonic, and the transverse and longitudinal components begin to display a rich behavior, that can be possibly thought as the combination of a gas-like and a solid-like contribution as proposed in \cite{PhysRevResearch.6.013206}. The correlated motion also appears at the level of the single particle displacement field. This field now presents geometric structures composed of vortex-like and string-like patterns and an increasing degree of heterogeneity with localized areas of large displacement (intense purple color) separated by more rigid regions can be observed (see panel (b) in Fig.\ref{fig:2})).

\begin{figure}[ht!]
    \centering
    \includegraphics[width=\linewidth]{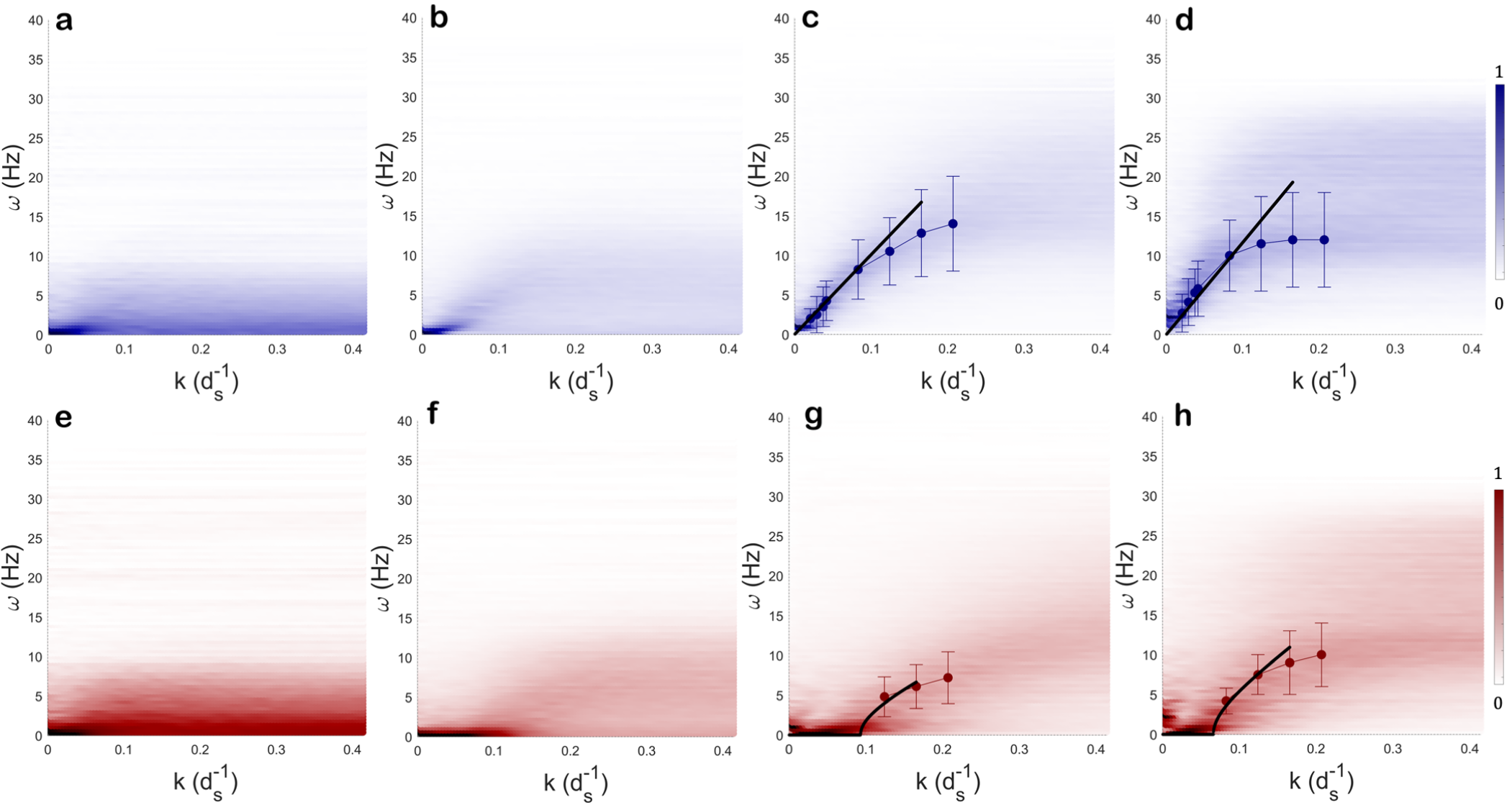}
    \caption{\textbf{Emergence of wave-vector gapped shear waves in liquid granular matter.} The longitudinal (\color{blue}blue\color{black}) and transverse (\color{red}red\color{black}) dynamical structure factors $S_{L,T} (\omega,k)$ for packing fraction $\phi=0.618$, $\phi=0.773$, $\phi=0.811$ and $\phi=0.822$ from \textbf{(a)} to \textbf{(d)} on top and from \textbf{(e)} to \textbf{(h)} on bottom. $d_s$ is the diameter of small particles. In panels \textbf{(c)} and \textbf{(d)}, the solid black lines show the fitting to the linear dispersion relation $\mathrm{Re}(\omega)=v_L k$. The fitted longitudinal speed of sound is $v_L = 101.2$ and $v_L = 116.9$ respectively in unit of $(d_s /s)$. The colored bullets indicate the peak position obtained by fitting $S_L (\omega,k)$ at fixed values of the wave vector $k$. In panels \textbf{(g)} and \textbf{(h)}, the numerical data are fitted with the $k$-gap formula $\mathrm{Re}(\omega)=\sqrt{v_T^2 (k^2-k_g^2)}$ where $k_g$ is the size of the $k$-gap. We obtain a $k$-gap of $0.093$ for $\phi=0.811$ \textbf{(g)} and $0.068$ for $\phi=0.822$ \textbf{(h)} in units of $d_s^{-1}$. The fitted transverse speed of sound is $v_T = 48.5$ and $v_T = 71.7$ respectively in unit of $(d_s /s)$. In all panels, the error bar indicates the linewidth of the corresponding peak in the dynamic structure factor. The dynamical structure factor is dimensionless and normalized by its maximum value, hence, the color bar goes from $0$ to $1$ in all panels.}
    \label{fig:3}
\end{figure}

Moving further to $\phi=0.811$ (panel (c) in Fig.\ref{fig:2}) results in the disappearance of the Lorentzian gas-like contribution to the VDOS at low frequency. The total VDOS increases monotonically with frequency up to a peak located around $\omega=15$ Hz, which corresponds approximately to the average pseudo-Van Hove energy of the emergent collective longitudinal and transverse excitations, as shown in Fig.\ref{fig:3}. The zero-frequency values of the total VDOS $g(0)$ exhibit a substantial decrease with an increase in the packing fraction $\phi$. This value correlates with the self-diffusion constant $D$, and a direct relation between $D$ and $g(0)$ can be derived for pure Langevin diffusion. This is consistent with the experimental data for the MSD presented in panel (c) of Fig.\ref{fig:1}. $g(0)$ is largely dominated by the transverse component, and $g_L(0)$ is almost zero at $\phi=0.811$. Additionally, the system becomes strongly anisotropic as the longitudinal and transverse VDOS are significantly different, and higher-energy modes appear in the spectrum up to a frequency of approximately $35$ Hz.
The maximum frequency mentioned is determined by the highest acquisition rate of the cameras we use to track the position of particles in our granular matter system, which is 40 Hz. This maximum frequency has no intrinsic physical significance and is unrelated to the driving frequency.

As the packing fraction of the system increases to $\phi=0.822$, the value of $g(0)$ becomes extremely small but still finite, indicating that the system is close to the solid phase but not quite there yet. This is later confirmed by the dispersion of shear waves. The longitudinal and transverse VDOS are linear in frequency up to about 8 Hz, as dictated by Debye's law in 2D, and also commonly found in bulk liquids \cite{doi:10.1073/pnas.2022303118,doi:10.1021/acs.jpclett.2c00297}. However, a sharper peak appears around 12 Hz, which is attributed to the flattening of the dispersion of the collective modes obtained from the dynamical structure factor (panels (d) and (h) in Fig.\ref{fig:3}). As a result, particle displacements become small, and granular particles move very little away from their initial positions. The dynamics and corresponding VDOS increasingly resemble those of a dense viscous liquid with enhanced solid-like elastic vibrations or a liquid near melting.

In summary, the analysis of the VDOS and the particle displacements reveal a continuous transition from a gas-like behavior typical of dilute liquids to a collective and viscoelastic motion characteristic of dense liquids by increasing the packing fraction. This perfectly aligns with the structural and dynamical transition between a gas-like liquid and a rigid liquid phase discussed in the previous section and displayed in Fig.\ref{fig:1}. The study of collective modes performed in Fig.\ref{fig:3} confirms that $\phi$ plays the role of the inverse temperature of classical thermal liquids (see Fig.\ref{fig:temp} and related discussion above). Indeed, the behavior of the VDOS shown in panels (a)-(d) in Fig.\ref{fig:2} is perfectly compatible with that found in liquids upon decreasing $T$ (see, for example, Fig.6 in \cite{PhysRevResearch.6.013206}).

\subsection*{Gapped shear waves}
To analyze the vibrational dynamics of our system, we consider the experimental dynamical structure factor $S_{L,T} (\boldsymbol{k},\omega)$ (details in Methods). Fig.\ref{fig:3} shows a color map of the obtained dispersion relations of the longitudinal and transverse parts for different packing fractions $\phi$ ranging from the gas-like phase $\phi=0.618$ to the dense liquid-like phase $\phi=0.822$. At the low packing fraction $\phi=0.618$, no distinct collective excitation is observed in both the longitudinal and transverse sectors (panels (a) and (e)). Instead, an incoherent diffusive signal, which extends from zero frequency to approximately $10$ Hz roughly independent of the wave vector, dominates the response. Moreover, the longitudinal and transverse components are almost identical, consistent with the emergent isotropy at low packing fractions and the results for the VDOS in panel (a) of Fig.\ref{fig:2}. This is also compatible with the structural analysis and the dynamical crossover observed in the VACF shown in Fig.\ref{fig:1}.

At packing fraction of $\phi=0.773$, panels (b) and (f), the system is in a dilute liquid phase. This phase displays only weak indications of collective motion, with the exception of a strong diffusive signal around $\omega=k=0$. There are also faint indications of collective modes in the longitudinal and transverse sectors. However, these excitations are strongly overdamped, which makes it difficult to identify their energy using wave-vector cuts precisely. Despite this, there appears to be a $k$-gap emerging around $k=0.08 d_s^{-1}$, where $d_s$ is the diameter of the small particles. This can be anticipated despite the challenges in identifying the energy of the excitations due to the broadening of the color map in panels (b) and (f).

As we move to higher packing fraction data, $\phi = 0.811, 0.822$ shown in panels (c)-(g) and (d)-(h), the fingerprints of liquid-like collective modes become more pronounced. By fitting the dynamical structure factor, we have extracted the frequency $\Omega(k)$ and the line-width $\Gamma(k)$ of the lowest collective modes in both the longitudinal and transverse sectors. Details regarding the fitting process can be found in the Methods section, and additional figures can be found in the Supplementary note 3. In the longitudinal sector, as shown in panels (c) and (d), we observe a mode that disperses linearly at large wavelengths, with $\Omega_L(k)=v_L k$, where $v_L$ is of the order $10^2$ in units of $d_s$/s. The speed increases slowly as the packing fraction increases, confirming that $\phi$ plays the role of inverse temperature in classical thermal liquids. At $k \approx 0.1 \, d_s^{-1}$, the dispersion bends down towards a constant pseudo-Van-hove plateau. Additionally, the longitudinal sound mode has a linewidth that increases with the wave vector $k$ and becomes overdamped for small wavelengths.

The transverse component of the dynamical structure factor in the dense liquid phase at a high packing fraction is shown in panels (g) and (h). Apart from a strong diffusive signal near the origin, $S_T(k,\omega)$ presents a distinctive $k$-gap dispersion for the transverse shear waves, which is confirmed by using cuts at constant $k$ (red symbols). The value of the $k$-gap is approximately $k_g=0.093$ for $\phi=0.811$ and $k_g=0.068$ for $\phi=0.822$ in units of $d_s^{-1}$. As the packing fraction $\phi$ increases, the value of the $k$-gap becomes smaller, which is equivalent to decreasing the temperature in classical liquid systems. We find that the frequency of the gapped shear waves is well fitted by the theoretical formula $\Omega_T(k)=\sqrt{v_T^2 (k^2-k_g^2)}$, with a transverse speed of sound of the order of $v_T = 48.5$ and $v_T = 71.7$ in the unit of $(d_s /s)$ for $\phi=0.811$ and $\phi=0.822$, respectively. The linewidth of the shear waves becomes larger with the wave vector $k$, similar to longitudinal waves.

\begin{table}[htb]
    \centering
    \begin{tabular}{|p{0.8cm}||p{1cm}|p{1cm}|p{1cm}|}
        \hline
        $\phi$ & $v_L/v_{\mathrm{th}}$ & $v_T/v_{\mathrm{th}}$ \\
        \hline 
        $0.811$ & $86.0$ & $41.2$  \\
        \hline
        $0.822$ & $121.2$ & $74.3$  \\
        \hline 
    \end{tabular}
    \caption{\textbf{Speed of sound.} The sound velocities obtained from the experimental data in units of the thermal velocity $v_{\mathrm{th}}$ derived from $Z(0)$ of all particles (see Fig.\ref{fig:temp}).}
    \label{speed}
\end{table}

In simple fluids with steep interactions, it has been found that $v_L \approx 10 v_{\mathrm{th}}$ and $v_T \approx 5 v_{\mathrm{th}}$ near freezing \cite{10.1063/5.0157945}, where $v_{\mathrm{th}}$ is the thermal velocity of the system. Using our experimental data, we have estimated the ratio between the sound velocities and the thermal velocity computed from $Z(0)$ (see Table \ref{speed}). We found that both ratios are one order of magnitude larger than in simple fluids. We speculate that this difference is a direct consequence of the activity of the system and of the strong deviations from local thermal equilibrium.

The lifetime of local connectivity, or $\tau_{LC}$, in classical liquids, is closely related to the Maxwell relaxation time, or $\tau$, as per a research study \cite{PhysRevLett.110.205504}. For a value of $\phi=0.822$, our analysis yields a value of $\tau_{LC}$ as 0.3 seconds, which is in proximity to the Maxwell relaxation time of 0.11 seconds, calculated from the size of $k$-gap (as shown in panel (d) of Fig.\ref{fig:3}). This similarity in values confirms the validity of our data analysis method for obtaining the dynamical structure factor. Also, it suggests that classical liquids in the supercritical state and granular fluids share similarities at the smallest particle level.

\begin{figure}[htb]
    \centering
    \includegraphics[width=0.8\linewidth]{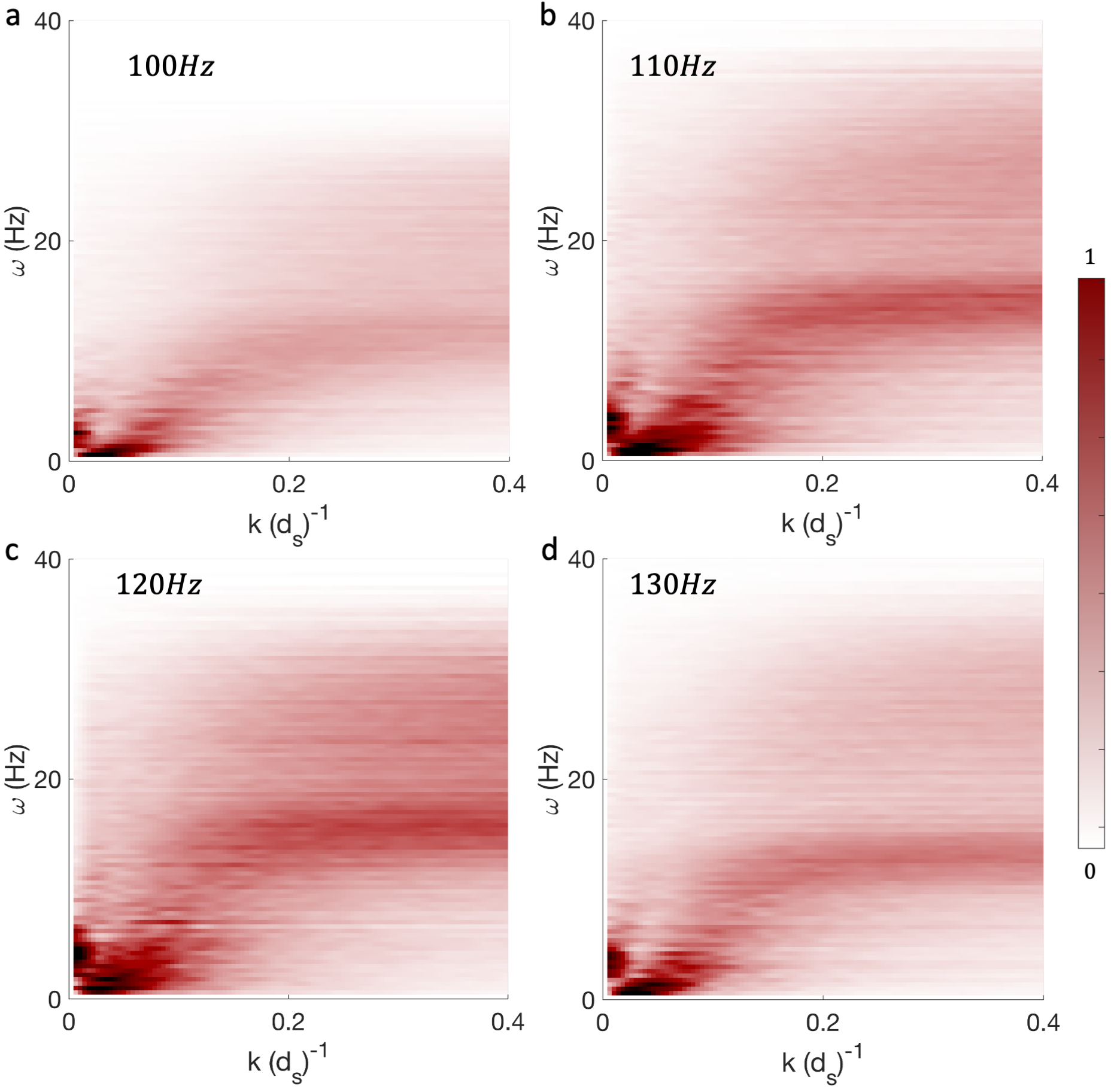}
    \caption{\textbf{Transverse dispersion relation under different driving frequency.} The transverse dynamical structure factors under different driving frequencies $f=100,110,120,130$ Hz from (a) to (d). The packing fraction is fixed to $\phi=0.822$, and the acceleration is fixed to $a=2.5g$.}
    \label{figdifferentfrequency}
\end{figure}

To understand how various parameters in the experimental setup is also necessary, in addition to the packing fraction $\phi$, influence the $k$-gap behavior. We conducted additional experiments to investigate this by increasing the driving frequency from 100 Hz to 130 Hz while maintaining the highest packing fraction of $\phi = 0.822$. We want to emphasize that the acceleration of the vibrator was held constant. This means that by changing the frequency $f$, the amplitude of the oscillations was also altered. In Fig.\ref{figdifferentfrequency}, we present the transverse spectrum for four different driving frequencies: $f = 100, 110, 120, 130$ Hz, with a constant packing fraction of $\phi = 0.822$ and acceleration $a = 2.5g$.

\begin{figure}[htb]
    \centering
    \includegraphics[width=0.8\linewidth]{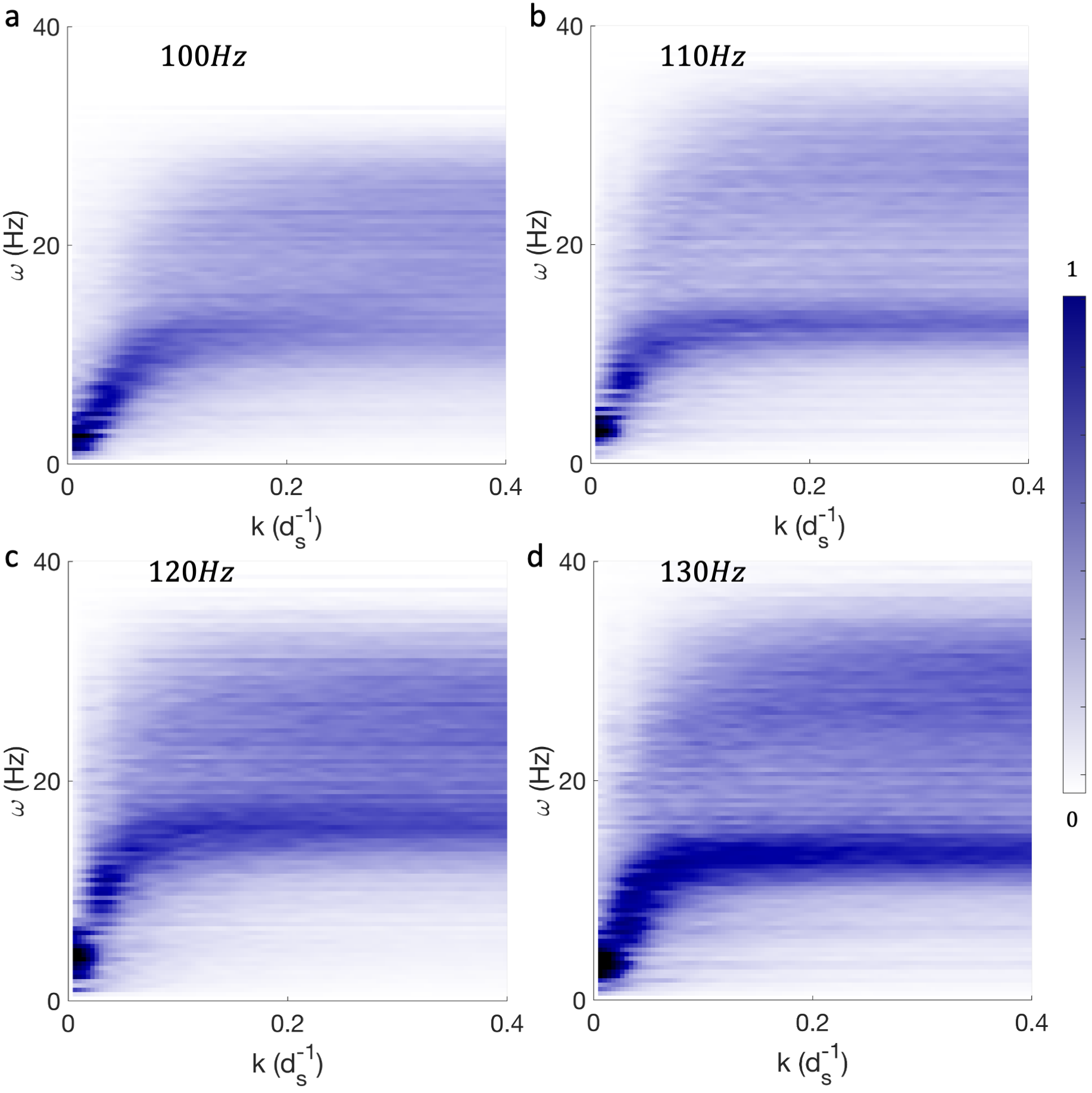}
    \caption{\textbf{Longitudinal dispersion relation under different driving frequency.} The longitudinal dynamical structure factor under different driving frequencies $f=100,110,120,130$ Hz from (a) to (d). The packing fraction is fixed to $\phi=0.822$ and the acceleration to $a=2.5g$.}
    \label{figdifferentfrequencyl}
\end{figure}

We observe that the position of the $k$-gap is largely independent of the driving frequency, at least in the range we considered. In contrast, the dispersion of shear waves is sensitive to the driving frequency and exhibits a complex dependence on the frequency $f$. A similar pattern is seen in the dispersion of longitudinal phonons, as illustrated in Fig.~\ref{figdifferentfrequencyl}. Further experimental investigation is necessary to thoroughly analyze how collective motion varies with driving frequency and amplitude.

Although we cannot conduct experiments with frictionless granular particles for direct comparison, we believe that interparticle friction is less significant than the interactions resulting from inelastic collisions between particles. Several pieces of evidence from our previous study of monodisperse particles \cite{Yangrui-PRE, Yangrui-NC} support this hypothesis.

First, we observed no correlations between individual particles' translational and rotational degrees of freedom. This suggests that particle rotation due to interparticle friction is largely independent of their translational motions. Second, we successfully explained the flocking behavior observed in our previous monodisperse system \cite{Yangrui-NC} by applying the theory of active Brownian particles \cite{CapriniPRL-2023}, which completely disregards interparticle friction, as the theory assumes. 

Third, we employed vertical driving instead of the more conventional shear driving, reducing interparticle friction's impact. Lastly, even in our bidisperse systems at the highest packing fraction, the system remains unjammed, indicating that the dominant interactions are due to interparticle collisions rather than contact forces associated with permanent contacts, as seen in jammed solids. Therefore, we anticipate that the results will be comparable to those of frictionless particles.

\section*{Conclusion}
Our study demonstrates that, despite significant differences in the size of composed particles and the absence of a classical thermodynamic description, granular matter on the Brownian vibrator exhibits various similarities with classical liquids in the supercritical region both in the collective and particle-level motions. These alike aspects include pair correlation functions, velocity autocorrelation, mean square displacement on particle levels, vibrational density of states, and dispersion relation of collective excitations. With an increase in the packing fraction $\phi$, distinct phases from gas-like to condensed states (liquid or solid) have been observed on both particle and collective motion levels, suggesting the role of $\phi$ as an effective inverse temperature variable, that is corroborated by the anti-correlation between the average kinetic energy and the packing fraction. However, it is needed to notice that the analogy of temperature is only vague and intuitive and is limited to our specific system. Furthermore, our work experimentally revealed the emergence of the $k$-gap, as predicted by viscoelasticity theory, in granular fluids. This provides a direct link between classical liquid and active granular matter and experimentally confirms the phenomenon of $k$-gap and its theoretical premises.

\section{Methods}
\subsection*{Experimental setup} 
Our experimental system comprises a horizontal layer of granular particles driven vertically by a sinusoidal oscillation with a fixed frequency $f$ and amplitude $A$ induced by an electromagnetic shaker.

Unless indicated otherwise, the frequency $f$ is set at 100 Hz, and the maximum acceleration $a$ is 2.5 times the gravitational acceleration $g=9.8 m/s^2$. The vibration amplitude $A$, defined as $a /(2 \pi f)^2$, is 0.062 mm, significantly smaller than the particle's vertical dimension (6 mm). Therefore, we can neglect the vertical displacement of a particle, treating the system as quasi-two-dimensional \cite{Yangrui-PRE}.

The upper left corner in Fig.\ref{fig:0} depicts a granular particle as a disk-shaped body with 12 alternately inclined supporting legs. The small particle's disk has a diameter $d$ of 16 mm and a thickness of 3 mm. The legs, with a height of 3 mm, are inclined inward by $18.4^o$, and alternately deviated from the mid-axis plane by $\pm38.5^o$.

For a single particle, the distributions of rotational velocity and translational velocity components $v_x$ and $v_y$ follow Gaussian distributions with negligible mean values compared to their standard deviations, typically less than ten percent. This indicates that the motion of a single particle is both random and isotropic, leading us to term the particle an Active Brownian Particle (ABP), closely mimicking conditions studied in theoretical investigations of active matter systems\cite{CapriniPRL-2023}.

To prevent crystallization, we utilize bi-disperse particles with a size ratio measured in terms of disk diameters of $1:1.4$ and a number ratio of $2:1$ for small and large particles. These parameters maintain an approximately equal area ratio between small and large particles across a range of packing fractions. The packing fraction $\phi$ is defined as the ratio between the area occupied by all particles and the confining area of the particle layer.

 These ratios are derived from jamming studies of binary disks, particularly the research conducted by O'hern et al. \cite{Ohern-PRL2002, Ohern-PRE2003}. Their simulations employed a size ratio of 1:1.4 because a large disk can have up to seven nearest neighbor small disks, while a small disk can have up to five nearest neighbor large disks. This configuration aligns with the theoretical concept of the 7-5 defect roles in the KTHNY theory \cite{Kardar}. 

In the original O'hern algorithm, a 1:1 number ratio was utilized. However, it was later discovered that a 2:1 number ratio between small and large disks, which corresponds to an area ratio of 1:1, yields even better results. A recent experimental study on the vibrational density of states in jammed 2D granular packing \cite{Zhangling-PRR-2021} analyzed how the density of states changed when the number ratio varied from a hexagonal crystal (1:0) to the most disordered binary mixture (2:1). The density of states for a system with a 2:1 ratio of small to large disks exhibited several features remarkably similar to those found in molecular glasses \cite{Zhangling-PRR-2021, Zhang-NC}.

The particles are placed on top of an aluminum plate (60 cm × 60 cm) and confined within a flower-shaped boundary designed to suppress creep particle motions along the boundary \cite{Yangrui-PRE}.
We initiate the experiment with all particles randomly and uniformly placed on the base plate. After applying vibration for two hours, we achieve an initial state of particle configuration. Subsequently, the particle layer undergoes continuous vertical vibration, while a Basler CCD camera (acA2040-180kc) records particle motion at 40 frames/s for at least an hour.

\subsection*{Displacement correlation matrix and dynamical structure factor} 

The displacement correlation matrix $\boldsymbol{C}$ is defined as,
\begin{equation}
    C_{i j}=\langle n(t)_i n(t)_j \rangle_t,
\end{equation}
where $n_i (t)$ is the displacement of $i$th degree of freedom at time $t$. The dynamical matrix can hence be calculated as,
\begin{equation}
    D_{i j}=\dfrac{\alpha\, C^{-1}_{i j}}{\sqrt{m_i m_j}},
\end{equation}
where $m_i$ is the mass of the $i$th degree of freedom, and $\alpha$ is a dimensionful parameter which will be later specified. Diagonalizing the matrix $\boldsymbol{D}$, one obtains the eigenvalues $\kappa_i$ and the eigenfrequencies $\omega_i = \sqrt{\kappa_i}$. The eigenvector fields $\boldsymbol{u}$ are then defined by solving the eigenvalue problem,
\begin{equation}
    \boldsymbol{D}\boldsymbol{u}= \omega^2 \boldsymbol{u}.
\end{equation}
Up to this point, the precise numerical values of the obtained eigenfrequencies $\omega_i$ are only determined up to an unknown energy scale $\alpha$. In athermal systems, such a scale cannot be determined using the temperature $T$, since the latter is not well defined. See \cite{C2SM07445A} for an extensive discussion on this issue. In order to fix the value of $\alpha$, we resorted to a more phenomenological approach. More precisely, we have rescaled all the eigenfrequencies by setting the value of the largest one to coincide with the highest observable frequency of our instrument, namely $40Hz$. In doing so, the corresponding eigenfrequencies have now the correct physical dimension. We validate a posteriori our hypothesis by noticing that the time scale obtained from $k$-gap $\tau$ is compatible with the average local connectivity time, as observed in classical thermal liquids \cite{PhysRevLett.110.205504}.

As an ulterior check of the validity of our criterion to fix $\alpha$, we compute the mean particle kinetic energy, $\langle E_{\text{kin}} \rangle \equiv m Z(0)$ (see Fig.\ref{fig:temp}). Following the literature \cite{PhysRevE.101.032903}, we can define an effective granular temperature $T_e$, and determine $\alpha$ using  $\alpha=k_B T_e=\langle E_{\text{kin}} \rangle $. By adopting this method, we find a normalization for $\alpha$ that is compatible, apart from an $\mathcal{O}(1)$ prefactor, with the previous estimates. This confirms the validity of our previous arguments. Finally, we emphasize that the value of $\alpha$, apart from fixing the physical dimension of our frequencies, affects only the values of the frequencies but not the qualitative physical trends discussed in the main text, as for example the shape of the VDOS or of the dispersion of collective excitations. At the same time, the value of the $k$-gap does not depend on the determination of $\alpha$.

To separate the longitudinal and transverse components properties, we Fourier transform the eigenvector fields
\begin{equation}
    F_{L,i} (\boldsymbol{k})=\boldsymbol{k} \cdot \sum_j \boldsymbol{u}_{i} (\boldsymbol{r}_j) e^{-i \boldsymbol{k} \cdot \boldsymbol{r}_j},
\end{equation}
and
\begin{equation}
    F_{Tz,i} (\boldsymbol{k})=\boldsymbol{z} \cdot \boldsymbol{k} \times \sum_j \boldsymbol{u}_{i} (\boldsymbol{r}_j) e^{-i \boldsymbol{k} \cdot \boldsymbol{r}_j},
\end{equation}
where the index $i$ indicates the eigenvector field corresponding to eigenfrequency $\omega_i$, the index $j$ indicates the $j$th particle, $\boldsymbol{r}_j$ is the equilibrium position of $j$th particle and $\boldsymbol{u}_{i,j}$ is the eigenvector field corresponding to frequency $\omega_i$ and position $\boldsymbol{r}_j$. $T,L$ stand respectively for transverse and longitudinal. Finally, $z$ is the spatial coordinate perpendicular to the 2D $(x,y)$ plane. The current correlation function can be obtained as,
\begin{equation}
    C_{T,L} (\boldsymbol{k},\omega)=\left| \sum_i F_{(T,L),i} (\boldsymbol{k}) \delta_{\omega,\omega_i} \right|^2 .
\end{equation}
The dynamical structure factor are given by,
\begin{equation}
    S_{L,T} (\boldsymbol{k},\omega) \propto \dfrac{1}{k^2} \left| \sum_i F_{(L,T),i} (\boldsymbol{k}) \delta_{\omega,\omega_i} \right|^2 .
\end{equation}

Finally, the vibrational density of states (VDOS) can be obtained directly by counting the distribution of eigenfrequencies $\omega_i$, which can be expressed as,
\begin{equation}
    g(\omega)=\sum_i \delta (\omega-\omega_i).
\end{equation}
Alternatively, one can integrate the current correlation function to obtain longitudinal and transverse components separately. The separated VDOS are given by,
\begin{equation}
    g_{L,T}(\omega) =\mathcal{N} \int  C_{L,T} (\boldsymbol{k},\omega) d \boldsymbol{k}
\end{equation}
where $g(\omega)=g_L(\omega)+g_T(\omega)$ and $\mathcal{N}$ is just a normalization factor to ensure that $\int g(\omega) d\omega=2N$ where $N$ is the number of particles. These two methods give same results for the total VDOS.
\subsection*{Data analysis}
The following fitting functions have been used to analyze the experimentally obtained dynamical structure factor (see additional figures in the Supplementary note 3). The transverse and longitudinal components of the dynamical structure factor are respectively fitted using
\begin{align}
  &  S_{T}(\omega,k) \propto \dfrac{\omega^2}{(\omega^2-\Omega_{T}^2)^2+ \omega^2 \Gamma_T^2} + \dfrac{1}{\pi}\dfrac{\gamma}{\omega^2 + \gamma^2}+S_{\text{loc}}(\omega,k),\\
  &S_{L}(\omega,k) \propto \dfrac{\omega^2}{(\omega^2-\Omega_{L}^2)^2+ \omega^2 \Gamma_L^2}.
\end{align}
The first terms are simply a damped harmonic oscillator with energy $\Omega_{L,T}(k)$ and linewidth $\Gamma_{L,T}(k)$. The second term is a quasi-elastic contribution modelled with a Lorentzian function. The dispersion relations $\Omega_{T,L}(k)$ are shown in Fig.\ref{fig:3} as colored bullets and the corresponding linewidths $\Gamma_{T,L}$ by the related errors bars. $S_{\text{loc}}(\omega,k)$ is a term representing possible quasi-localized low-energy modes \cite{Hu2022}, and it is not relevant for the present discussion. More details on the fitting procedure and the raw data for the dynamical structure factors can be found in the Supplementary note 3.

\section{Data availability}
The datasets generated and analysed during the current study are available upon reasonable request by contacting the corresponding authors. 

\section{Acknowledgments}
CJ and MB acknowledge the support of the Shanghai Municipal Science and Technology Major Project (Grant No.2019SHZDZX01). MB acknowledges the support of the sponsorship from the Yangyang Development Fund.
ZZ, YC, and JZ acknowledge the support of the NSFC (No. 11974238 and No. 12274291) and the Shanghai Municipal Education Commission Innovation Program under No. 2021-01-07-00-02-E00138. ZZ, YC, and JZ also acknowledge the support from the Shanghai Jiao Tong University Student Innovation Center.

\section{Author contributions statement}
M.~B. and J.~Z. conceived the idea of this work and supervised it. Z.~Z and Y.~C performed the experiments. C.~J. performed the data analysis and the theoretical modeling with the help of M.~B. All authors contributed to the writing of the manuscript.

\section{Competing interests}
Jie Zhang is an Editorial Board Member for Communications Physics, but was not involved in the editorial review of, or the decision to publish this article. All the other authors declare no competing interests.

\section{Preprints}
A preprint of this article is available as, arXiv:2403.08285.



\clearpage

\renewcommand\thefigure{S\arabic{figure}}    
\setcounter{figure}{0} 
\renewcommand{\theequation}{S\arabic{equation}}
\setcounter{equation}{0}
\renewcommand{\thesection}{S\arabic{section}} 
\setcounter{section}{0}

\section*{\Huge Supplementary notes}

\section*{Supplementary note 1: Debye normalized density of states}
\begin{figure}[htb]
    \centering
    \includegraphics[width=\linewidth]{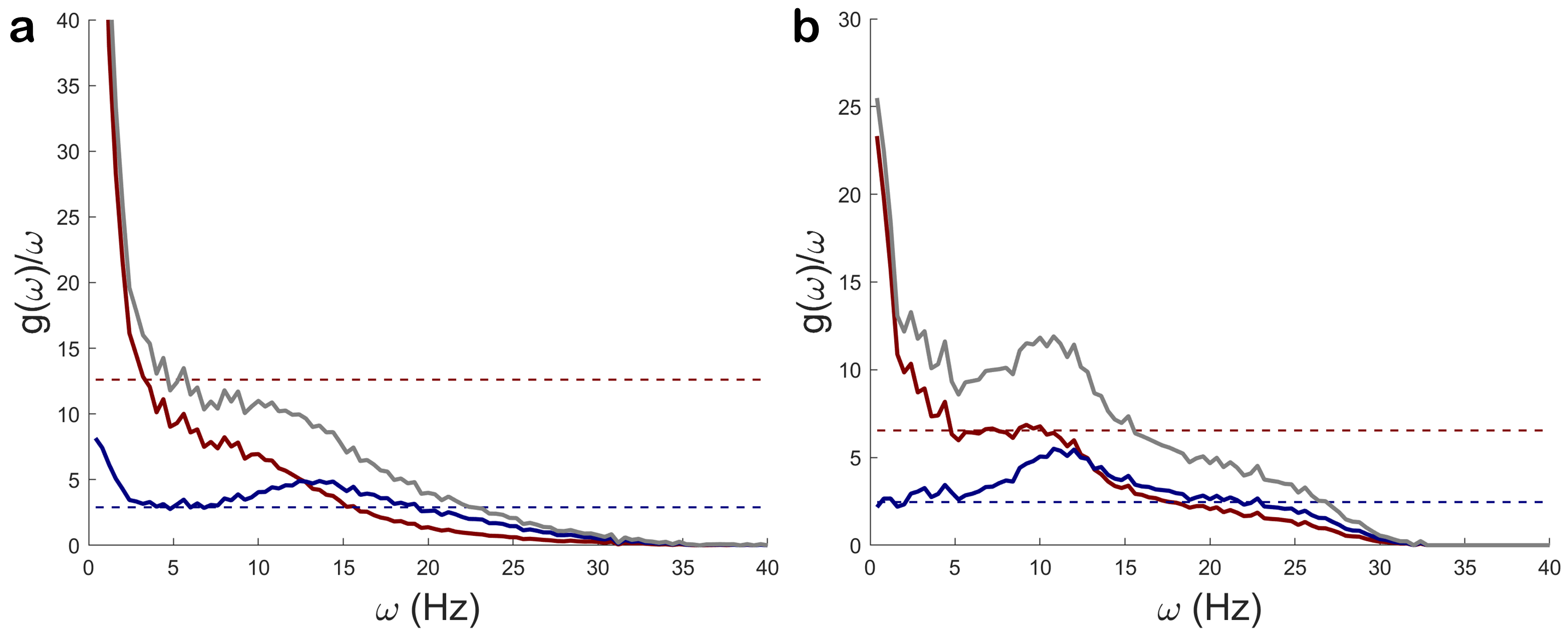}
    \caption{The reduced VDOS for packing fraction $\phi=0.811$ \textbf{(a)} and $\phi=0.822$ \textbf{(b)}. The blue lines indicate the longitudinal component, while the red and the gray indicate respectively the transverse and total ones. The dashed lines show the theoretical predictions from the Debye model as explained in the text.}
    \label{rdos}
\end{figure}

For convenience, the vibrational density of states for solid materials is often normalized by the prediction from Debye's theory, which holds only for ideal crystals with long-range order. In two dimensions, Debye's theory predicts that the VDOS at low frequency is given by
\begin{equation}
    g_{\text{Debye}}(\omega)= \frac{L^2}{2\pi \bar{v}^2}\,\omega \qquad \text{with} \qquad \frac{2}{\bar{v}^2}=\frac{1}{v_T^2}+\frac{1}{v_L^2}.
\end{equation} 
The reduced VDOS $g(\omega)/\omega$ is presented for the two highest values of the packing fraction in Fig.\ref{rdos}.

For the highest packing fraction, $\phi=0.822$, the longitudinal component of the reduced VDOS approaches a constant plateau at low frequency, implying the validity of the Debye model. On the contrary, the transverse component follows Debye's law only between approximately $10$ Hz to $5$ Hz and then strongly deviates from it below such a frequency. This manifests a residual quasi-elastic diffusive contribution and a finite value of $g_T(\omega)$ at zero frequency. By extracting the speeds of longitudinal and transverse sound from the dispersion relation of the collective modes, we have verified that the ratio of the two plateau values in the reduced VDOS (dashed horizontal lines in Fig.\ref{rdos}) is compatible with the theoretical prediction of Debye theory, namely $v_T^2/v_L^2$.

The situation becomes more complex by decreasing the packing fraction, as shown in the left panel of Fig.\ref{rdos} for $\phi=0.811$. There, Debye's model no longer provides a good description of the experimental data. We can only observe a small plateau in the longitudinal component of the reduced VDOS around $5$ Hz, which is then modified at lower frequencies. In general, by increasing the system's fluidity or equivalently decreasing its packing fraction, we no longer expect the Debye model to be a good representation of the collective dynamics.

\section*{Supplementary note 2: Static structure factor}
\begin{figure}
    \centering
    \includegraphics[width=0.5\linewidth]{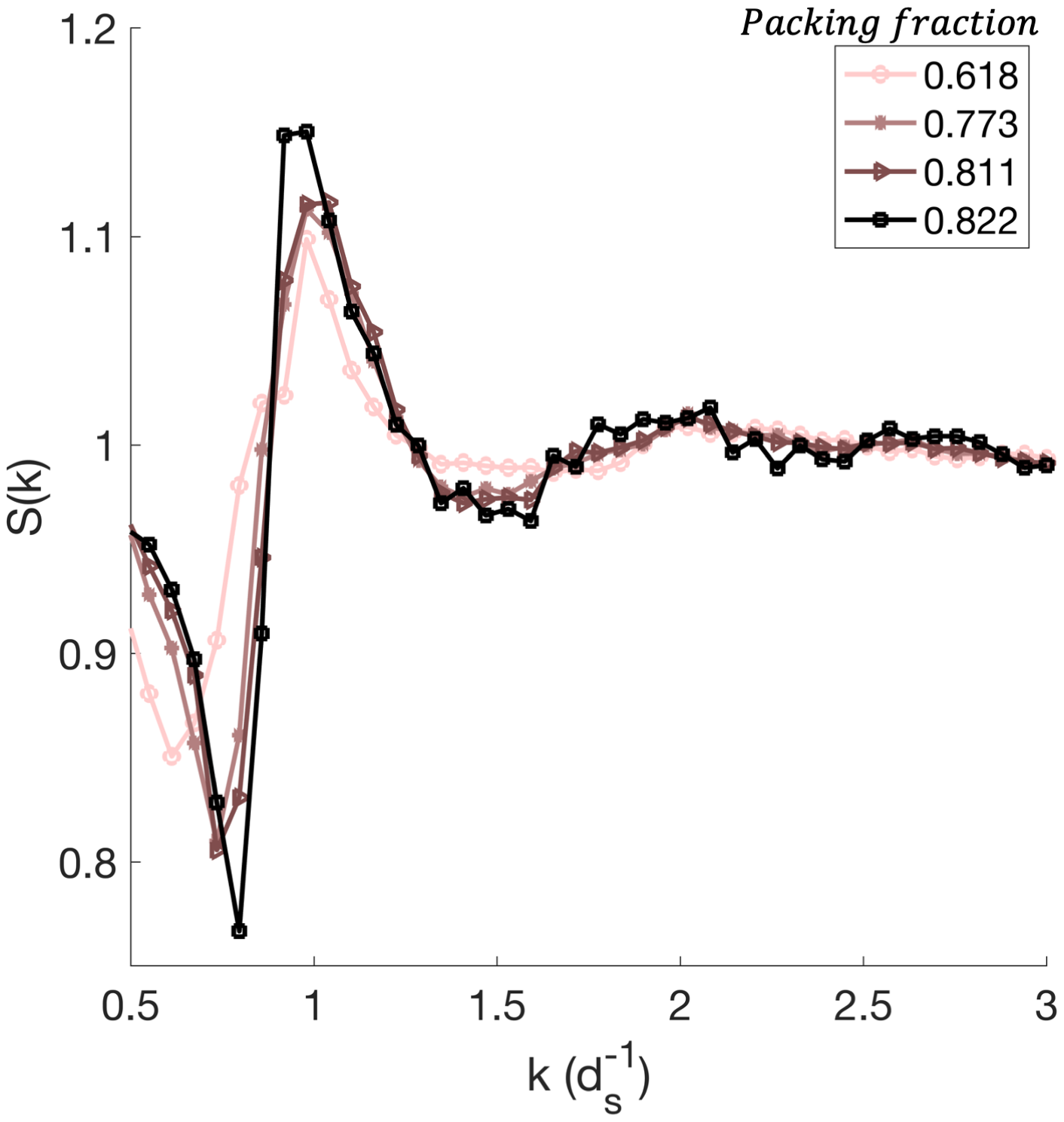}
    \caption{The experimental static structure factor for different values of the packing fraction $\phi$.}
    \label{figsk}
\end{figure}

We have obtained the experimental structure factor $S(k)$ for four different packing fractions to provide further experimental evidence for a connection between the dynamical Frenkel line and structural changes. The results are shown in Fig.~\ref{figsk}. From there, a gradual disappearance of the medium-range structural order is evident by decreasing $\phi$ from the maximal value $\phi=0.822$ to a low value $\phi=0.618$. In particular, by reducing $\phi$, we notice that (I) the intensity of the first peak around $k\approx d_s^{-1}$ diminishes, (II) the depth of the valley around $k\approx 1.5 d_s^{-1}$ becomes smaller and (III) the higher order peaks gradually disappear. These results imply a gradual loss of medium-range order from the dense liquid phase (large $\phi$) to the gas-like dilute phase (small $\phi$). 

\section*{Supplementary note 3: Dynamical structure factor}
We provide more details about the analysis of the experimental dynamical structure factor. The longitudinal dynamical structure factor is fitted using a single damped harmonic oscillator,
\begin{equation}
    S_{L}(\omega,k) \propto \dfrac{\omega^2}{(\omega^2-\Omega_{L}^2)^2+ \omega^2 \Gamma_L^2},
\end{equation}
with $\Omega_L(k),\Gamma(L)(k)$ respectively the energy and the line-width of the corresponding mode.
The transverse dynamical structure factor is fitted using three contributions,
\begin{equation}
    S_{T}(\omega,k) \propto \dfrac{\omega^2}{(\omega^2-\Omega_{T}^2)^2+ \omega^2 \Gamma_T^2} + \dfrac{1}{\pi}\dfrac{\gamma}{\omega^2 + \gamma^2}+S_{loc}(\omega,k),
\end{equation}
with
\begin{equation}
    S_{loc}(\omega,k)=\dfrac{1}{\omega \sigma \sqrt{2 \pi}} \mathrm{exp}\left( -\dfrac{(\mathrm{ln} \omega-\mu)^2}{2 \sigma^2} \right).
\end{equation}
The equation consists of three terms. The first term corresponds to a damped harmonic oscillator contribution for the longitudinal part. The second term is a quasi-elastic Lorentzian line shape, which takes into account the low-energy diffusive dynamics. Lastly, the third term represents a log-normal function that considers the possible presence of quasi-localized low-energy modes as in glasses.

\section*{Supplementary note 4: Displacement field power spectrum}
We have performed the Fourier transform on the displacement field to observe its evolution over time. Fig. \ref{figftdisplacement} presents the results for different packing fractions. For low packing fractions (a) and (b), all Fourier transform components concentrate strongly and stably in the low wave vector region after a long observation time (800s). This indicates that the particles do not return to their initial position. However, the features differ for high packing fractions, as shown in panels (c) and (d). First, the Fourier components are unstable, especially the zigzag structures at $k \approx 0$. This suggests that the particles may still return to their initial position. Secondly, the Fourier transform components concentrate at a wave vector smaller than $k$-gap. This correlation can be interpreted as follows: the modes with $k>k_g$ are solid-like, meaning that the particles only vibrate around their equilibrium position, and therefore these modes do not contribute to the displacement at a late time. On the other hand, the modes with $k<k_g$ are gas- or liquid-like, meaning that the particles can diffuse away from their position and never return. Therefore, only these modes contribute to the long-time displacement.

\section*{Supplementary note 5: Lifetime of local atomic connectivity}
We calculate the lifetime of local atomic connectivity, $\tau_{LC}$, which represents the average time it takes for atoms to lose or gain one nearest neighbor, changing the local atomic connectivity through a local configurational excitation. The Voronoi cell of the particle at time $t=t_0$ defines its nearest neighbors.
The schematic diagram of Fig.\ref{neighbor_map} indicates that any two particles that share an edge of a Voronoi cell are considered neighbors, as explained in detail in \cite{Voronoi-PRL}. In doing so, the atomic bonds of a particle with its neighbors can be defined. After that, we can detect the formation or breaking of atomic bonds for $t>t_0$ by identifying the moments when the number of neighbors changes by a unity.
For each atom, $t_{LC}$ is the time to change one neighbor such that $N_{c}(t_{0}+t_{LC})=N_{c}(t_{0})\pm1$. The lifetime of local atomic connectivity is then defined as the average of $t_{LC}$, which is denoted as $\tau_{LC}=\langle t_{LC} \rangle$.
The distribution of $t_{LC}$ for different packing fractions is shown in Fig.\ref{Pnormal(t_lc)}. $\tau_{LC}$ values for the $\phi$ values 0.618, 0.773, 0.811, and 0.822 are 0.308, 0.609, 0.507, and 0.300s, respectively.

\begin{figure}[htbp]
    \centering
    \includegraphics[width=0.8\linewidth]{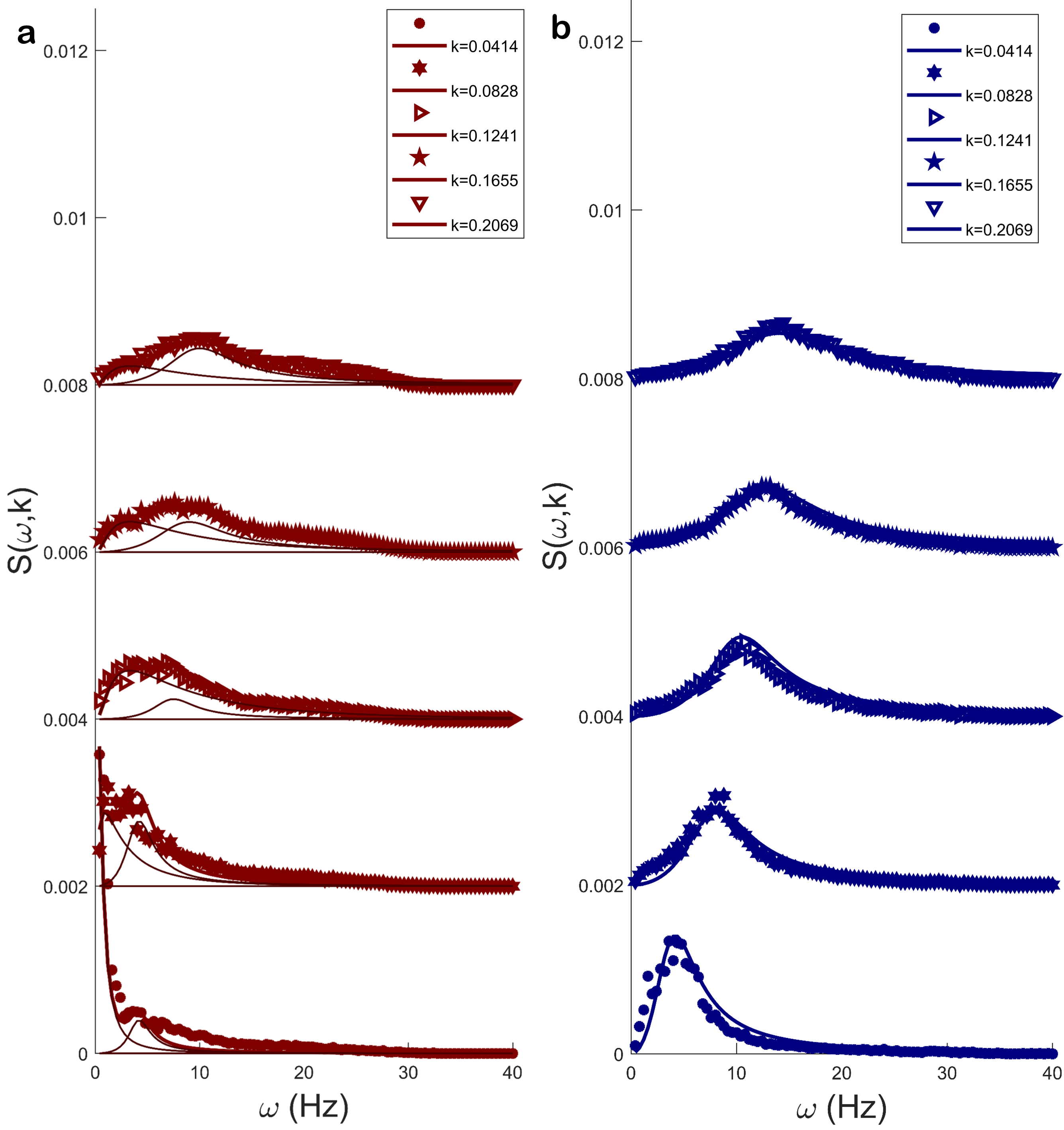}
    \caption{Transverse (red,\textbf{a}) and longitudinal (blue,\textbf{b}) dynamical structure factor at different wave vector for $\phi=0.822$. The thinner lines in transverse panel show the three terms in the fitting function.}
    \label{dsflines787}
\end{figure}

\begin{figure}[htbp]
    \centering
    \includegraphics[width=0.8\linewidth]{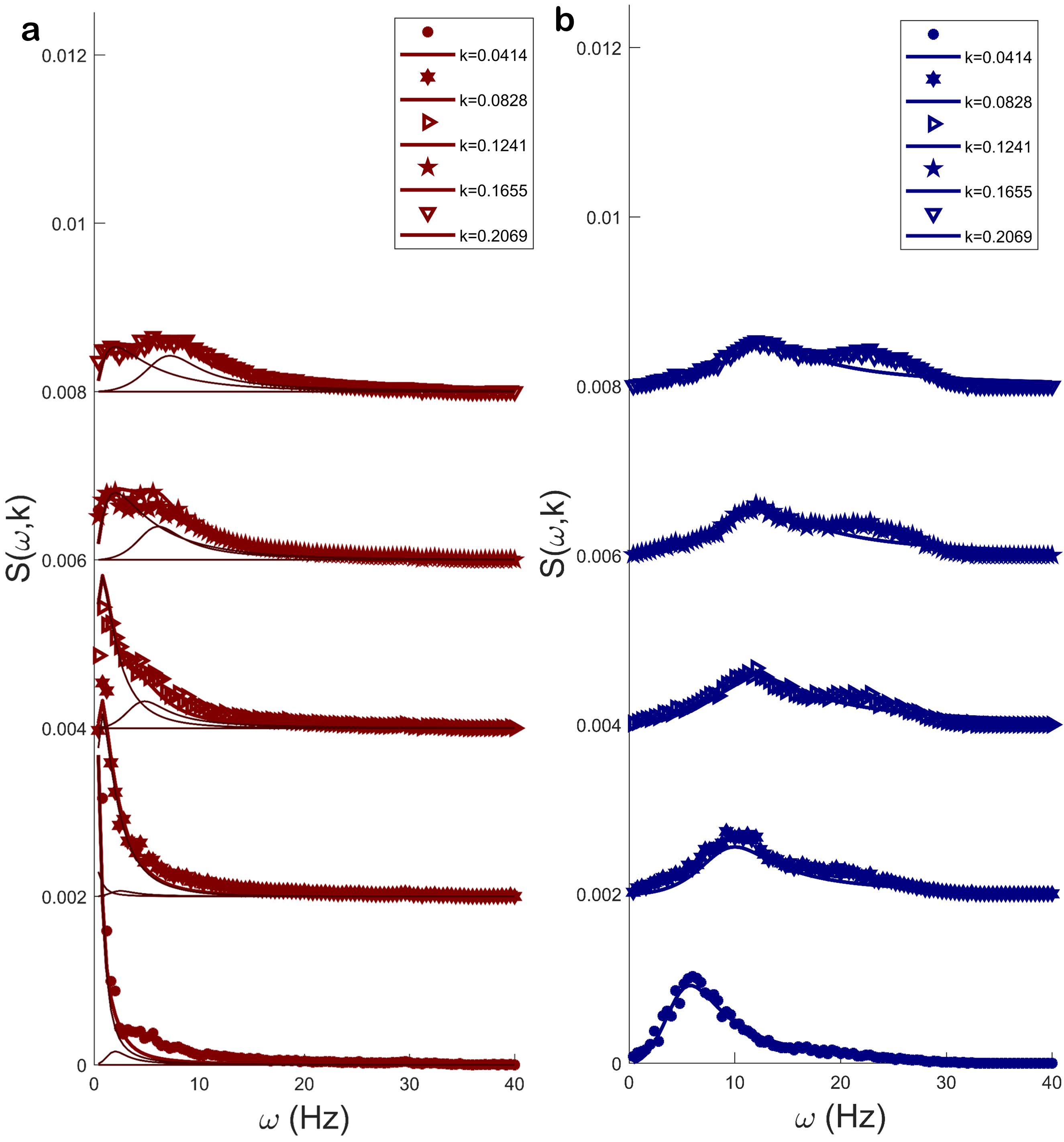}
    \caption{Transverse (red,\textbf{a}) and longitudinal (blue,\textbf{b}) dynamical structure factor at different wave vector for $\phi=0.811$. The thinner lines in transverse panel show the three terms in the fitting function.}
    \label{dsflines798}
\end{figure}

\begin{figure}[htbp]
    \centering
    \includegraphics[width=\linewidth]{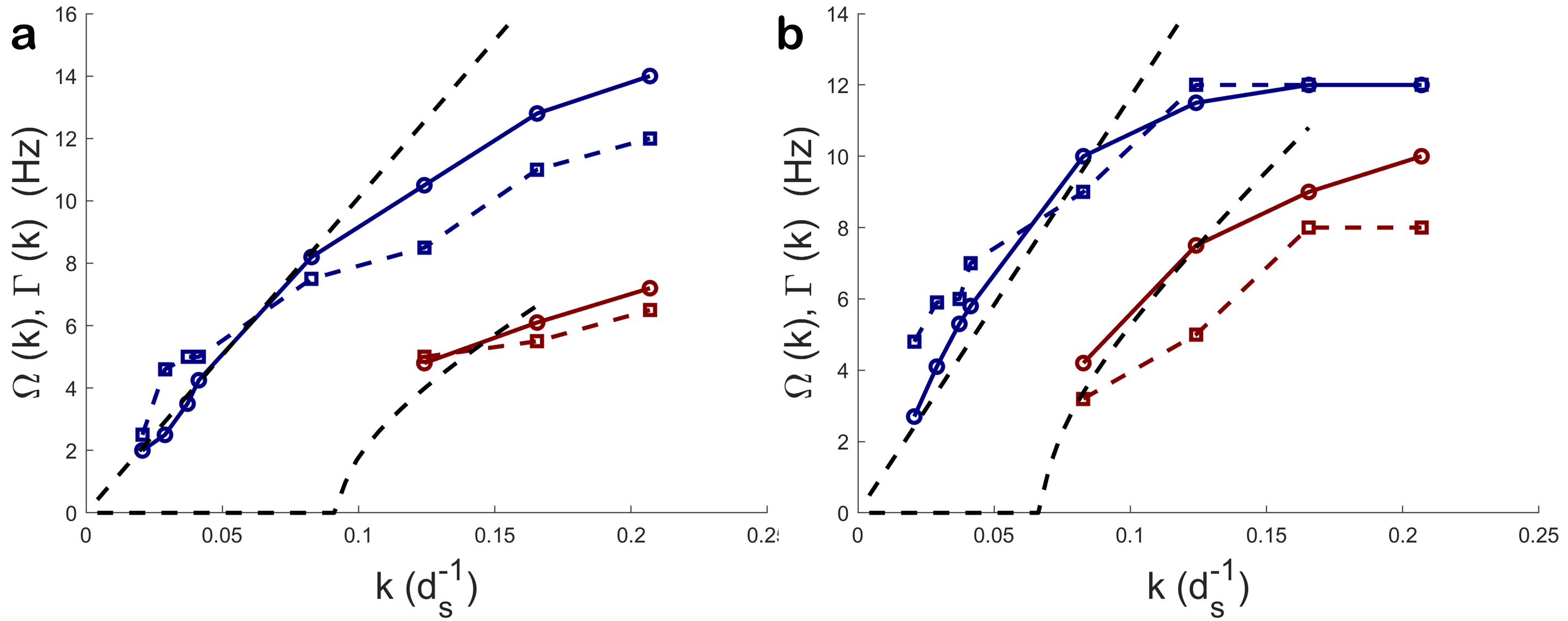}
    \caption{The fitting parameters $\Omega$ (circle) and $\Gamma$ (square) for longitudinal (blue) and transverse (red) dynamical structure factor and for packing fraction $\phi=0.811$ \textbf{(a)} and $\phi=0.822$ \textbf{(b)}.}
    \label{fitting parameters}
\end{figure}

\begin{figure*}[htbp]
    \centering
    \includegraphics[width=\linewidth]{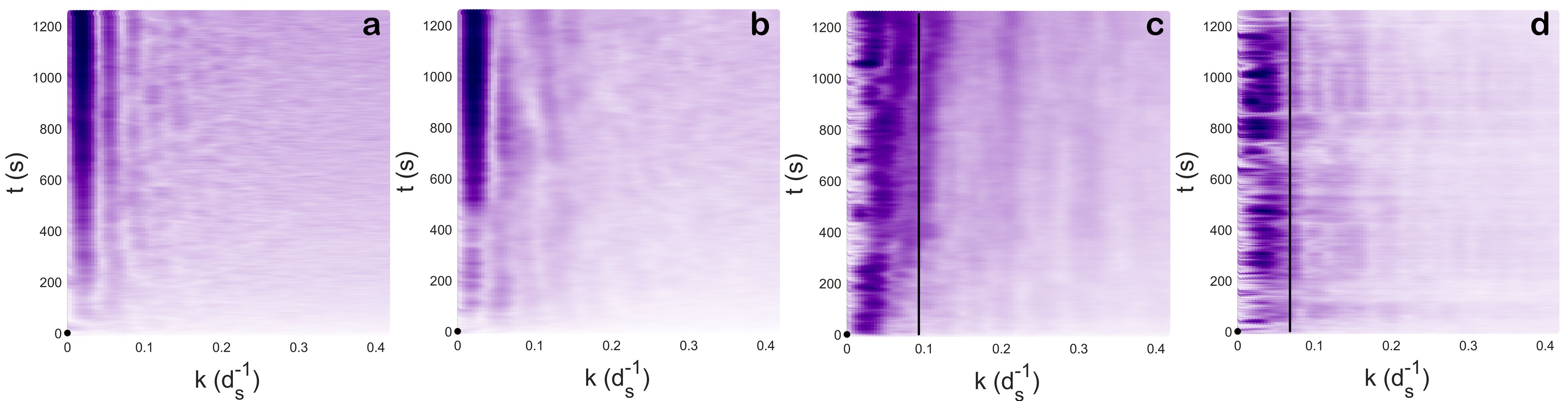}
    \caption{Fourier transform on displacement field, $\boldsymbol{D}(\boldsymbol{k})=\mathcal{F}[\boldsymbol{D}(\boldsymbol{x})]$, for packing fraction $\phi=0.618$, $\phi=0.773$, $\phi=0.811$ and $\phi=0.822$ from \textbf{(a)} to \textbf{(d)} respectively. Since our system have rotational symmetry, only $x$ component $D_x (\boldsymbol{x})$ are plot. The darker color indicate stronger Fourier transform signal. The red lines in panels (c) and (d) show the position of $k$-gap.}
    \label{figftdisplacement}
\end{figure*}

\begin{figure}[htbp]
    \centering
    \includegraphics[width=0.7\linewidth]{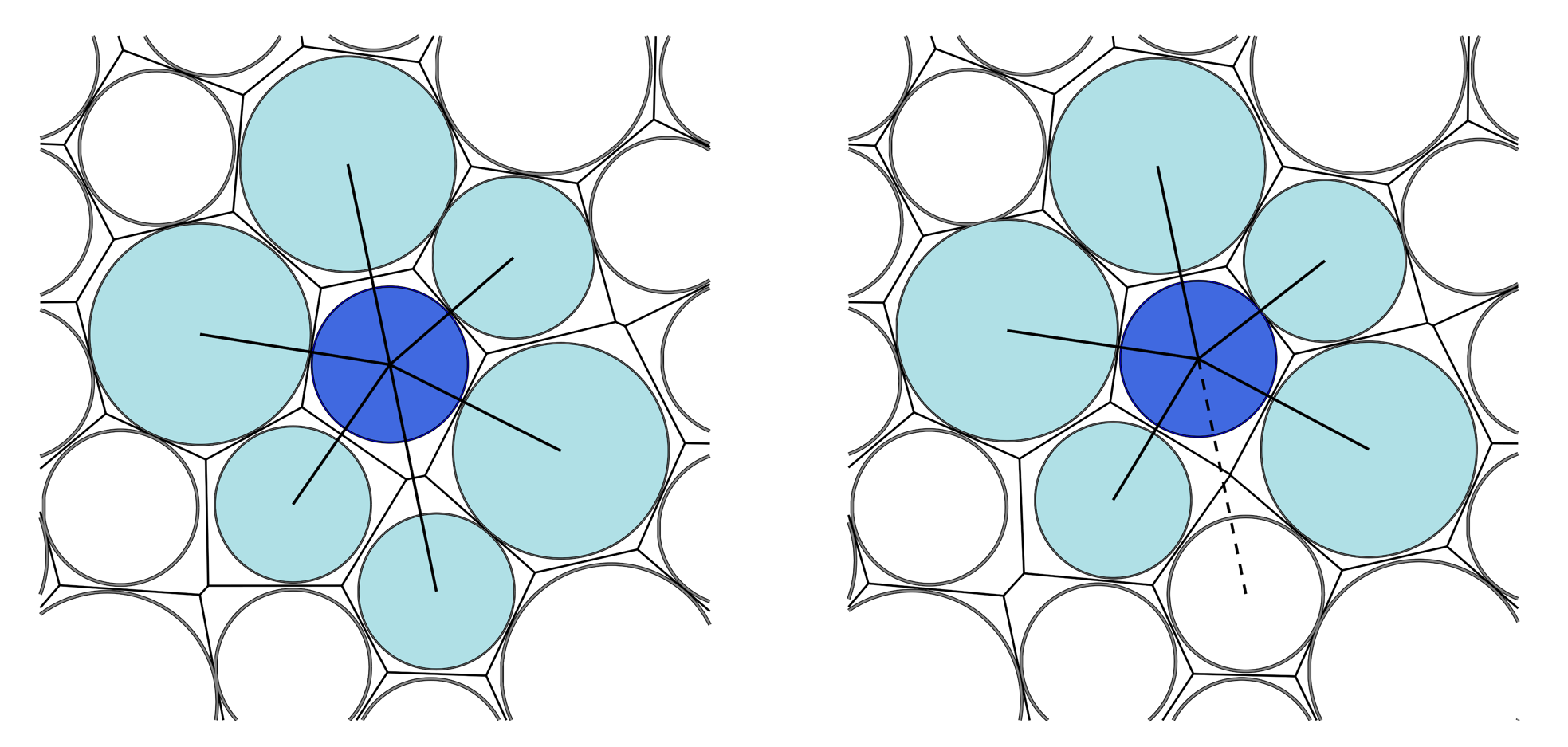}
    \caption{Schematic diagram for an atom to change one neighbor.}
    \label{neighbor_map}
\end{figure}

\begin{figure*}[htbp]
    \centering
    \includegraphics[width=\linewidth]{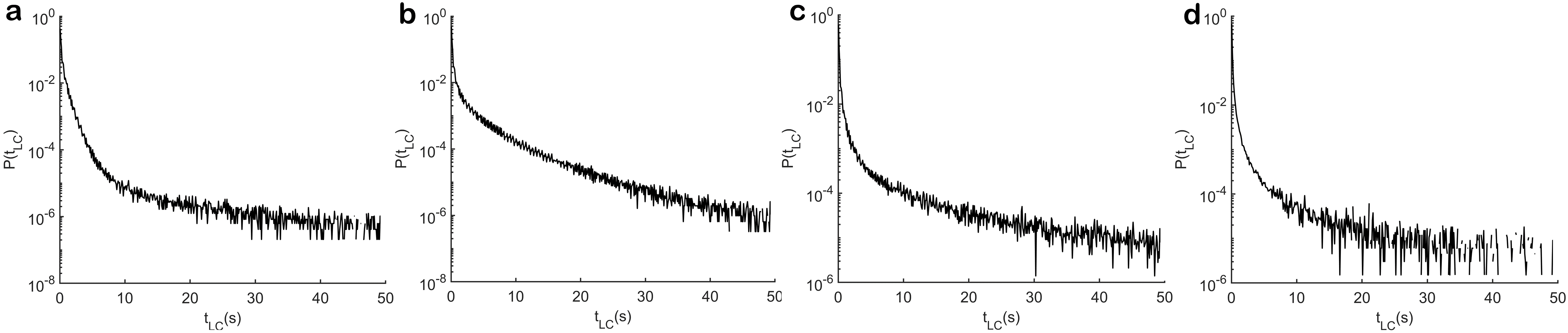}
    \caption{Distribution of $t_{LC}$ for packing fraction $\phi=0.618$, $\phi=0.773$, $\phi=0.811$ and $\phi=0.822$ from (a) to (d). $\tau_{LC}=0.308,0.609,0.507,0.300s$ for each packing fraction respectively.}
    \label{Pnormal(t_lc)}
\end{figure*}

\section*{Supplementary note 6: Dynamics of a single active particle}

\begin{figure}[htb]
    \centering
    \includegraphics[width=0.65\linewidth]{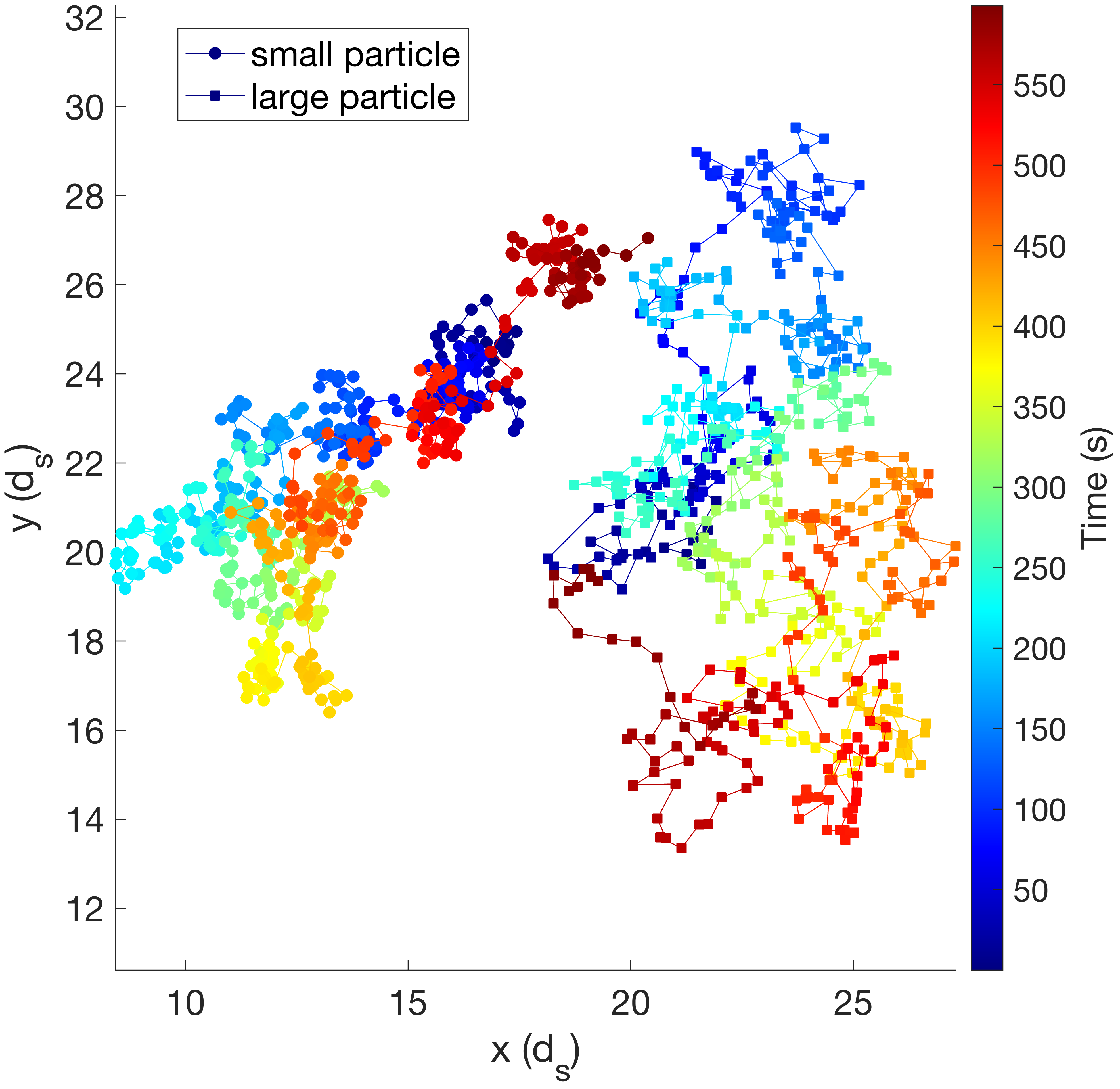}
    \caption{Typical trajectories of small and large particles. The color indicates the time evolution within an observation window of $600$ seconds.}
    \label{path}
\end{figure}

In Fig. \ref{path}, we show two typical trajectories for one small particle and one large particle under an observation time of $600$ s (from blue to red color). 

\begin{figure}[htb]
    \centering
    \includegraphics[width=0.45\linewidth]{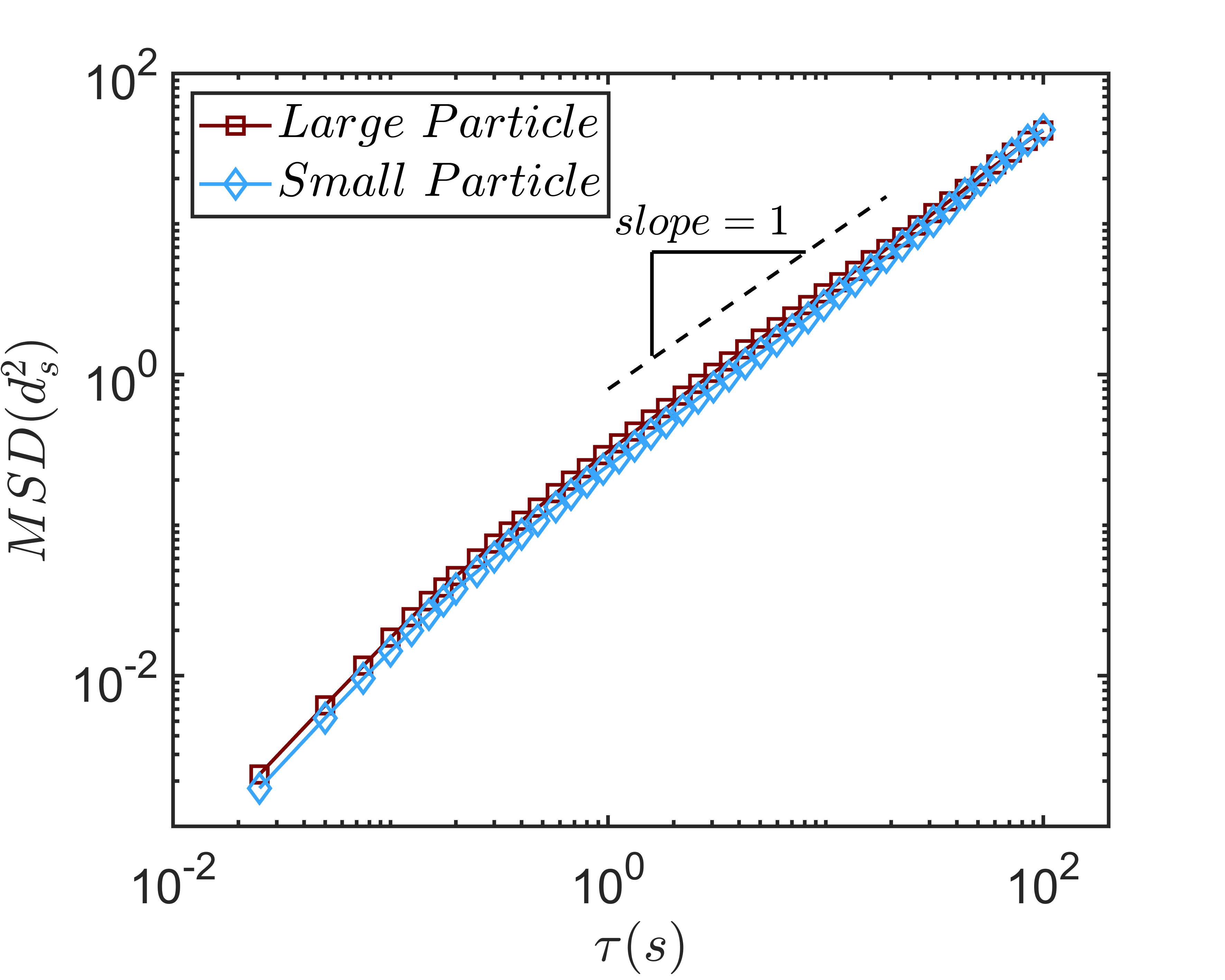}
    \includegraphics[width=0.45\linewidth]{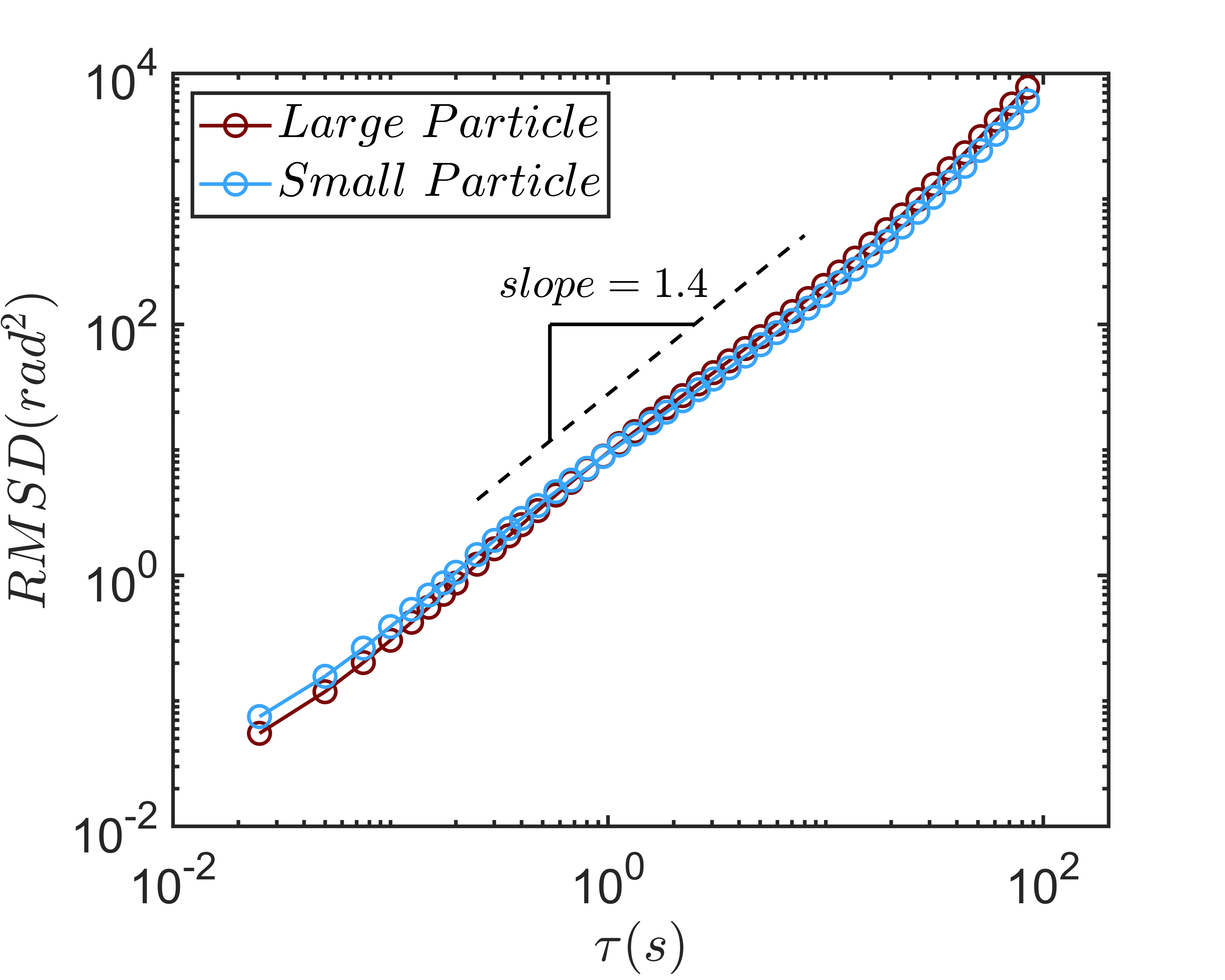}    
    \caption{The mean square displacement MSD (left) and rotational mean square displacement RMSD (right) based on single particle experiment for both small and large particles.}
    \label{msdall}
\end{figure}

The MSD and RMSD for small and large particles are, respectively, shown in panels (left) and (right) of Fig.\ref{msdall}. The left panel demonstrates that the MSD curves of both small and large particles are almost identical, exhibiting both short-time quasi-ballistic and long-time diffusive motions. The quasiballistic motion observed at short time scales corresponds to a persistent time of approximately $\tau_p \approx 0.25$  seconds for both small and large particles. The diffusion process begins for $\tau\ge\tau_p$ on the MSD curves, as illustrated in Fig.\ref{msdall}(left). From the long-time diffusive motions, we derive their translational diffusion constants $D \approx 0.45 d_s^2$/s for both small and large particles. 

The right panel illustrates that the RMSD curves for both small and large particles are nearly indistinguishable, indicating weakly super-diffusive motions. This is evidenced by a slope close to 1.4, as shown by the triangle on the log-log plot. The observed super-diffusive behavior arises from the technical challenges associated with manufacturing perfectly rotationally symmetric particles. This behavior can be quantified using the fitting function $\text{RMSD}=D_{r,{\ast}} t^{\gamma}$, with $\ast$ representing `l' and `s', yielding the constants $D_{r,l} = 9.12$, $D_{r,s} = 9.33$, $\gamma_l = 1.43$, and $\gamma_s = 1.35$, where the subscripts `l' and `s' refer to `large' and `small' particles, respectively.

The single-particle dynamics clarify the active nature of our experimental setup.

\end{document}